\documentclass[fleqn,10pt]{wlscirep}
\usepackage[utf8]{inputenc}
\usepackage[T1]{fontenc}
\usepackage{float}
\usepackage{physics}
\DeclareMathAlphabet{\mathcal}{OMS}{cmsy}{m}{n}
\usepackage[normalem]{ulem}
\begin{document}
\title{
 Easy-plane anisotropy of magnetic fluctuations in UTe$_2$ }
\author[1,2,+]{Zehao Wang}
\author[3,+]{Ewan Scott}
\author[4,5]{David W. Tam}    
\author[6]{Philippe Bourges}
\author[7]{Keke Feng}
\author[7]{Tyler W. Wannamaker}
\author[4]{Paul Steffens}
\author[4,8]{Arno Hiess}
\author[9]{Ryan E. Baumbach}
\author[7]{M. Brian Maple}
\author[3,10,*]{Michal P. Kwasigroch}
\author[1,2,**]{Pengcheng Dai}

\date{\today}

\affil[1]{Department of Physics and Astronomy, Rice University, Houston, TX 77005, USA}
\affil[2]{Rice Laboratory for Emergent Magnetic Materials and Smalley-Curl Institute, Rice University, Houston, TX 77005, USA}
\affil[3]{Department of Mathematics, University College London, Gordon St., London WC1H 0AY, United Kingdom}
\affil[4]{Institut Laue Langevin, 71 avenue des Martyrs, CS 20156, 38042 GRENOBLE Cedex 9, France}
\affil[5]{Department of Applied Physics, KTH Royal Institute of Technology, SE-106 91 Stockholm, Sweden}
\affil[6]{Université Paris-Saclay, CNRS, CEA, Laboratoire Léon Brillouin, 91191, Gif-sur-Yvette, France}
\affil[7]{Department of Physics, University of California San Diego, La Jolla, CA 92093, USA}
\affil[8]{European Spallation Source ERIC, P.O. Box 176, 22100 Lund, Sweden}
\affil[9]{Department of Physics, University of California Santa Cruz, Santa Cruz, CA 95064, USA}
\affil[10]{Trinity College, Cambridge, CB2 1TQ, United Kingdom}

\affil[*]{mpk32@cam.ac.uk}
\affil[**]{pdai@rice.edu}
\affil[+]{these authors contributed equally to this work}


\begin{abstract}
Superconducting UTe$_2$, which has an orthorhombic crystal structure, has generated much interest because the large directional anisotropy along the lattice $a$, $b$, $c$ directions in magnetic susceptibility, upper critical fields, and Knight shift suggests spin-triplet Cooper pairing. If spin fluctuations bind the Cooper pair, one needs to determine their anisotropy in spin space to characterize the superconducting state. 
Here we use polarised and unpolarised inelastic neutron scattering (INS) experiments to show that antiferromagnetic (AFM) spin fluctuations at the $Y$ point are highly isotropic within the $ac$-plane but have no observable magnitude along the $b$-axis. 
Our results rule out the assumed Ising anisotropy and constrain the possible pairing channels mediated by spin fluctuations.  AFM exchange, e.g., between the uranium ladders, is no longer pair forming in the triplet channel and the $\mathbf{d}$ vector along $b$ is strongly favoured.
We capture the experimental observations within a microscopic framework that encompasses the dual nature of uranium's $5f$ electrons.
We include both itinerant and localized $5f$-moments in a tight-binding model fitted  independently to quantum oscillations data.
In congruence with INS measurements, the exchange between localized moments generically consists of, by far the strongest, FM intra-dimer interaction between nearest neighbours along $c$, maxima close to the $Y$ and $T$ points of the Brillouin zone, mediated by $5p_y$ and $6d_{3z^2-r^2}$ orbitals respectively, and magnetically decoupled $ab$-planes. We find that anisotropy and strong dimerization of the moments can 
lead to gapped excitations. In combination with Landau damping, these can 
produce an overdamped response that is observed in the vicinity of the $Y$ point. We suggest that enhancement of the response and its observed near $O(2)$ symmetry could both be driven by proximity of the localized moments to a quantum critical point where the dimer excitation gap closes and gives place to incommensurate helical order. Analysis of the linearized gap equation for a nearest-neighbour pairing interaction favours the $B_{1u}$ triplet with a dominant $\mathbf{d}$-vector component along $b$, although the possibility of spin singlet pairing cannot be ruled out.

\end{abstract}

\flushbottom
\maketitle

\thispagestyle{empty}

\section{Introduction}

In conventional Bardeen–Cooper–Schrieffer (BCS) superconductors, electrons form coherent spin-singlet ($S=0$) Cooper pairs via electron-lattice coupling and open an isotropic superconducting gap at the Fermi level below the superconducting transition temperature $T_c$ \cite{PhysRev.108.1175}. 
 For unconventional superconductors such as copper \cite{WOS:000349190300029}, iron \cite{RevModPhys.87.855}, heavy fermions \cite{Stewart03042017,10.3389/femat.2022.944873}, or nickel-based materials \cite{10.1093/nsr/nwaf373}, superconductivity appears near antiferromagnetic (AFM) ordered phases, Cooper pairs are believed to be spin-singlets but mediated by spin fluctuations at finite momentum transfer $\mathbf{Q}$ in contrast to conventional phonon-induced superconductors \cite{RevModPhys.84.1383}. In the magnetic ordered phase of unconventional superconductors such as BaFe$_2$As$_2$, the presence of spin-orbit coupling (SOC) locks the AFM moment direction along the $a$-axis of the orthorhombic structure \cite{RevModPhys.87.855}, and induces different gaps in spin excitations along the $M_a$ ($a$), $M_b$ ($b$), and $M_c$ ($c$) axis directions, denoted as $\Delta_a$, $\Delta_b$, and $\Delta_c$, respectively, at the AFM ordering wave vector $\mathbf{Q}$ with $\Delta_a>\Delta_b>\Delta_c$   
 \cite{PhysRevB.86.060410,PhysRevX.3.041036}.
 At sufficiently low energy, $M_c>M_b>M_a$ with $M_a$ and $M_b$ vanishing below their respective thresholds to induce spin space anisotropy at $\mathbf{Q}$ at $T\ll T_N$ \cite{PhysRevX.3.041036}. Near $T_N$, $M_a\gg M_b=M_c$ before gapping out gradually  below $T_N$ \cite{Liu2020UniaxialPressure}.
By contrast, the uniform static susceptibility of detwinned BaFe$_2$As$_2$ below $T_N$ has approximately $\chi_b>\chi_c>\chi_a$ with $\chi_b-\chi_a$ increasing like an order parameter below $T_N$ \cite{He2017BaFe2As2Susceptibility}. Therefore, SOC can generate strongly momentum-dependent anisotropy, so the softest polarisation at AFM $\mathbf{Q}$ does not have to correspond to the largest magnetic response at $\mathbf{Q}=0$.
 When electrons and holes are doped into BaFe$_2$As$_2$ to induce superconductivity, low-energy spin space anisotropy persists into the superconducting state \cite{PhysRevLett.111.107006,PhysRevB.85.214516,PhysRevB.90.140502,PhysRevB.94.214516}, where the longitudinal spin excitations ($M_a$ and $M_c$ for electron- and hole-doped iron pnictides, respectively\cite{PhysRevB.90.140502,PhysRevB.94.214516}), likely the Anderson-Higgs mode \cite{PhysRev.110.827,annurev:/content/journals/10.1146/annurev-conmatphys-031214-014350}, are coupled to superconductivity \cite{PhysRevB.90.140502,PhysRevB.94.214516}. Similarly,
neutron polarization analysis reveals a $c$-axis polarized resonance at at AFM $\mathbf{Q}$ in nematic ordered FeSe \cite{PhysRevX.7.021025}.

\begin{figure}[H]
    \centering
    \includegraphics[width=\linewidth]{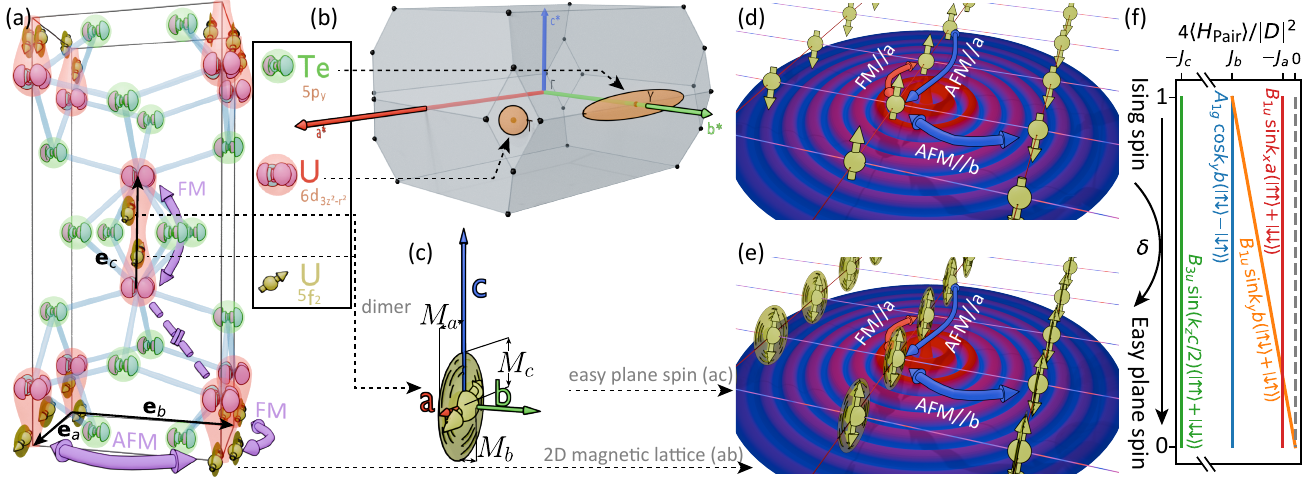}
    \caption{\textbf{Model of UTe$_2$. (a)} Crystal structure of UTe$_2$. The localized $5f_2$-moments (highlighted in yellow) belonging to the two U ions that constitute the dimer are strongly coupled by FM exchange. Pink arrows show the RKKY exchange between dimers. $6d_{3z^2-r^2}$ orbitals of the U are highlighted in red and $5p_y$ orbitals of Te-2 in green. The RKKY exchange between dimers belonging to different $ab$-planes is negligible. Definitions of $\mathbf{e}_a$, $\mathbf{e}_b$, and $\mathbf{e}_c$ for the theory part are marked by three black arrows. The axes of crystals is shown in (c). \textbf{(b)} 1st Brillouin zone (BZ) of UTe$_2$. The $6d_{3z^2-r^2}$-mediated RKKY leads to the spin excitations at the $T$ point, and the $5p_y$- mediated RKKY leads to the signal at the $Y$ point. \textbf{(c)} Spin fluctuations
    along $a$- ($M_a$), $b$- ($M_b$), and $c$-axis ($M_c$) directions.
    The fluctuating $5f_2$-moments are pinned to the $ac$-plane by the anisotropy of the intra-dimer FM exchange. \textbf{ (d, e)} Long-range RKKY exchange between the dimer moments in the $ab$-plane with either Ising (d) or easy-plane (e) anisotropy. The nearest-neighbour FM interaction along $a$ is partially frustrated by the next-nearest-neighbour AFM exchange. \textbf{(f)} As anisotropy changes from Ising (d) to easy $ac$-plane (e), the energy of the AFM correlated $B_{1u}$ triplet state vanishes -- the AFM exchange along $b$ is no longer pair forming in the triplet channel. The $B_{1u}$ and $B_{3u}$ states with dominant $\mathbf{d}$ vectors along $b$, that are driven by FM exchange, survive in the limit of easy $ac$-plane anisotropy.}`%
    \label{fig:crystal}
\end{figure}

Contrary to spin-singlet superconductors discussed above, orthorhombic lattice UTe$_2$ (Fig. \ref{fig:crystal}a) has been considered a promising candidate for spin-triplet ($S=1$) superconductor due in part to the large upper critical field ($H_{c2}$) exceeding the Pauli-limit, the uniform magnetic susceptibility along the $a$-axis suggestive of a nearby ferromagnetic (FM) instability, and the Knight shift from nuclear magnetic resonance (NMR) measurements \cite{doi:10.1126/science.aav8645,Aoki_2022,Matsumura2025}. 
However, INS experiments only found AFM spin fluctuations at finite momentum transfer $\mathbf{Q}$, and there is no evidence for FM spin fluctuations at $\mathbf{Q}=0$ \cite{PhysRevLett.125.237003,PhysRevB.104.L100409,Butch2022UTe2}. A map of reciprocal space reveals that the dominant spin excitations are centered at the symmetry point $Y$ in the middle of the edge of the Brillouin zone boundary at $\mathbf{Q}=(0,0.577,0)$ r.l.u. of the 2D rectangular lattice of U dimers, which has FM component along the $a$-axis and AFM component along the $b$-axis (Fig. \ref{fig:crystal}a, b) \cite{Duan2021,doi:10.7566/JPSJ.90.113706}.

The coexistence of FM and AFM interactions in UT$_2$ is similar to that of BaFe$_2$As$_2$, where SOC induces a clear spin-space anisotropy that couples to superconductivity \cite{PhysRevLett.111.107006,PhysRevB.85.214516,PhysRevB.90.140502,PhysRevB.94.214516}. 
Since UTe$_2$ is expected to have a much larger SOC compared to iron-based superconductors, it is important to determine the spin-space anisotropy $M_a$, $M_b$, and $M_c$
in UTe$_2$ at $\mathbf{Q}=(0,0.577,0)$ r.l.u and its role in the electron pairing mechanism.
Here we use polarised INS to show that spin excitations in UTe$_2$ are highly isotropic within the $a$-$c$ plane ($M_a\approx M_c$), but exhibit no observable excitations along the $b$ axis ($M_b\approx 0$, Fig. \ref{fig:crystal}c). These results rule out $a$-axis Ising antiferromagnetism as an origin of the possible spin-triplet superconductivity and suggest that any possible spin-triplet pairing must involve $c$ components of U-U dimers.
In the following, we first describe our unpolarised and polarised INS experimental results and then 
our theoretical model to compare with experiments.

\section{Unpolarised neutron scattering}
\begin{figure}[H]
    \centering
    \includegraphics[width=0.5\linewidth]{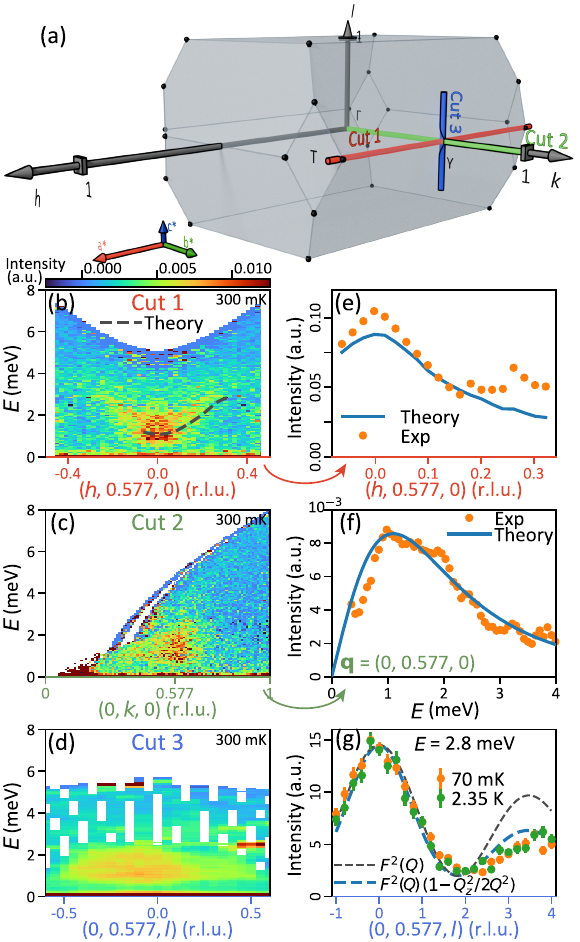}
    \caption{\textbf{Unpolarised neutron scattering. (a)} The reciprocal space of UTe$_2$. The dark blue brick is the 1st BZ. The red, green, and blue lines intersect at $(0,0.577,0)$ and represent the $x$-axis of (b-d). \textbf{(b-d)} Three cuts in the reciprocal space around the ${\bf {Q}}=(0, 0.577, 0)$ point as marked in panel (a) with scattering intensity against energy and momentum at $T=300$ mK. Those data combine the data of $E_i=3.32$ meV and $E_i=12$ meV, and the latter is multiplied by 5.47 to compensate for the difference in neutron flux of different $E_i$. The integration of panel (b) is $\Delta k=\pm0.05$ r.l.u. and $\Delta l=\pm 0.1$ r.l.u.. The integration of panel (c) is $\Delta h=\pm0.1$ r.l.u. and $\Delta l=\pm 0.1$ r.l.u.. The integration of panel (d) is $\Delta h=\pm0.1$ r.l.u. and $\Delta k=\pm 0.1$ r.l.u.. Panel (b) has been symmetrized by the mirror reflection of $[0,k,l]$ plane. The theoretically fitted dispersion is shown as a dashed line, which is fitted with raw data without combining $E_i=12$ meV and without symmetrization. \textbf{(e)} Comparison between the experimentally measured scattering intensity (orange dots) along the dispersion highlighted in panel (b) and the theoretical fit (blue line). \textbf{(f)} Comparison between the experimentally measured intensity (orange dots) and the theoretical fit (blue line) along the energy cut at ${\bf {Q}}=(0, 0.577, 0)$. The fit is done using Eq. \ref{eq:res}, and the position of the peak and maximum intensity of the fit are shown in panels (b) and (e). The fits are performed independently for each wave vector. \textbf{(g)} Triple-axis unpolarised data at $E=2.8$ meV along cut-3 direction at $T=70$ mK, and 2.35 K, together with the magnetic form factor $F^2(\mathbf{Q})$. A better fit is obtained if we assume equal contributions from fluctuations along $a$ and $c$, which leads to an additional scale factor of $1-Q_z^2/2Q^2$ in the theoretical prediction.}
    \label{fig:unpolarised}
\end{figure}

We first describe unpolarised neutron scattering experiments where 
some of the data were published in a previous work \cite{Duan2021}. 
Figure \ref{fig:unpolarised}a shows the locations of $Y$ and $T$ 
points in the three-dimensional (3D) 1st BZ of UTe$_2$. The dispersions of spin excitations along
the orthorhombic $a$, $b$, $c$ directions are consistent with previous work \cite{Halloran2025}. Figures \ref{fig:unpolarised}b-d show the cuts along these three axes as depicted by Fig. \ref{fig:unpolarised}a at 300 mK \cite{Duan2021}. The 2D maps in Fig. \ref{fig:unpolarised}b-d are taken at 300 mK. We combine data of $E_i=3.32$ meV and 5.47 times the data of $E_i=12$ meV for better illustration. Data in Fig. \ref{fig:unpolarised}b are symmetrized by a mirror plane at $[0,h,l]$. 
Inspection of the Figures reveal that spin excitations in UTe$_2$
are centered around the $Y$ point and disperse differently along the
three orthorhombic directions. To understand the dispersions, we use
 the raw data of $E_i=3.32$ meV only without any symmetric operations.
The black dashed line in Fig. \ref{fig:unpolarised}b and solid blue line in Fig. \ref{fig:unpolarised}e are our calculated dispersions along the $[h,0.577,0]$ direction (see the Theoretical model section). 
The orange data points in Fig. \ref{fig:unpolarised}e and Fig. \ref{fig:unpolarised}f are cuts along the $[h,0.577,0]$ direction 
and at $\mathbf{Q}=(0,0.577,0)$
in Fig. \ref{fig:unpolarised}b, respectively. 

The wave vector ${\bf {Q}}=(0, 0.577, 0)$, or the $Y$ point,
is the strangest point of magnetic fluctuations. Clear dispersive spin excitations are seen along the $a$-axis (Fig. \ref{fig:unpolarised}b). Along the $b$-axis, the dispersion is almost vertical (Fig. \ref{fig:unpolarised}c), indicating a strong coupling along the $b$-axis. Along the $c$-axis, the excitations are broad and diffusive (Fig. \ref{fig:unpolarised}d). Figure \ref{fig:unpolarised}g shows a separate unpolarised neutron scattering measurement using a triple-axis spectrometer along the same direction as Cut-3 does (Fig. \ref{fig:unpolarised}a). Constant energy scans with $E=2.8$ meV above (green dot) and below (orange dot) the superconducting temperature match the trend of the magnetic form factor of the U-dimer with projection factors (see Fig. \ref{fig:unpolarised}g), where we assume the magnetic fluctuations lie within the $ac$-plane with equal value. Using the Ising-type magnetism structure and comparing the simulation among the assumptions of magnetic static moments parallel to the $a$-axis, $b$-axis, or $c$-axis, it was found that the $a$-axis static magnetism fits the INS data best \cite{PhysRevB.104.L100409}. Since our neutron polarisation analysis reveal that spin excitations have components
along the $a$- and $c$-axes, and these components are equal ($M_a\approx M_c$ and $M_b\approx 0$, Fig. \ref{fig:polarised data}c), INS scans along the ${\bf {Q}}=(0, 0.577, l)$ direction, which probe spin excitations
perpendicular to ${\bf {Q}}$, have an extra projection factor. In Fig. \ref{fig:unpolarised}g, the gray dashed line represents the assumption of $a$-axis magnetism, and the blue dashed line represents the assumption of an equal weight of $a$- and $c$- components of magnetic fluctuations as suggested by the data. The fitting of other models can be found in the supplementary material. We believe the $ac$-plane magnetic fluctuation fits it best. The reason for the slower decay of the magnetic form factor in the theory calculation could be the deformation of the electron orbits due the U-dimerisation.

\section{Polarised neutron scattering}
\begin{figure}[H]
    \centering
    \includegraphics[width=\linewidth]{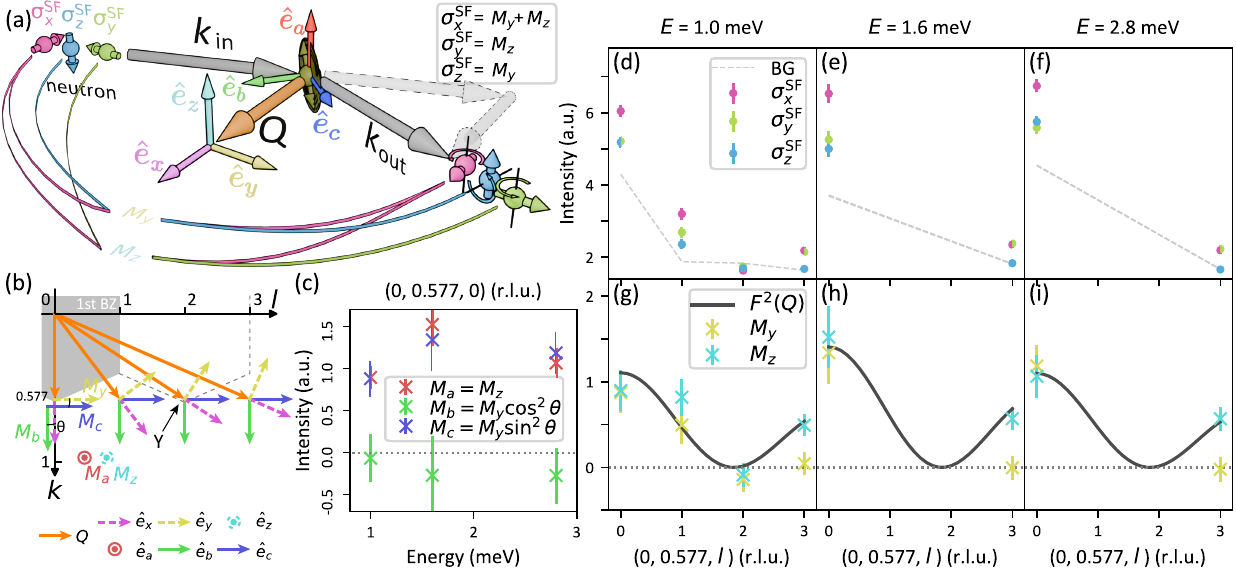}
    \caption{\textbf{Polarised neutron experiment within the $[0,k,l]$ plane at $T=70$ mK. (a)}, Scattering geometry. $\{\hat{e}_a,\hat{e}_b,\hat{e}_c\}$ and $\{\hat{e}_x,\hat{e}_y,\hat{e}_z\}$ represent the axes of the crystal and the axes of neutron scattering, respectively, where $\hat{e}_x$ is parallel to momentum transfer ${\bf Q}$, and $\hat{e}_z$ is parallel to $\hat{e}_a$ in our experiment. The balls and their arrows of different colors represent neutrons and their spins in different scattering channels. The experiment was taken near the $Y$ point. \textbf{(b)} Projection relationship between the axes of the crystal and the neutron scattering axes at different ${\bf Q}$ of the following data. $\theta$ is the angle between $\hat{e}_y$ and $\hat{e}_b$.    \textbf{(c)} Estimated spin excitations along the three axes of the crystal. \textbf{(d-f)} The raw $\sigma^{\rm SF}_{x,y,z}$ at $E = 1.0$, 1.6, and 2.8 meV as a function of
    $\mathbf{Q}=(0, 0.577, l)$. Estimated nonmagnetic scattering is drawn by a gray dashed line (see supplementary). \textbf{(g-i)} Calculated spin excitations along the neutron axes at $E = 1.0$, 1.6, and 2.8 meV. The magnetic form factor of U dimer is drawn as black lines.}
    \label{fig:polarised data}
\end{figure}

To determine spin space anisotropy in UTe$_2$, we carried out neutron polarisation analysis using the setup shown in Fig. \ref{fig:polarised data}a.  Three unit axes, $\{\hat{e}_a,\hat{e}_b,\hat{e}_c\}$, are defined parallel to the three orthorhombic axes of the crystal, respectively.  
The co-aligned single crystals in the $ab$-plane (the supplementary Fig. S1a) are aligned in the $[0,k,0]\times [0,0,l]$ horizontal scattering plane. We adopt a conventional coordinate system for polarised neutron studies with the $x$-axis along ${\bf {Q}}$, $y$-axis orthogonal to ${\bf {Q}}$ within the $[0,k,l]$ scattering plane, and $z$-axis perpendicular to the scattering plane (Fig. \ref{fig:polarised data}a). By aligning the incident neutron polarisation to be parallel to the $x$-, $y$-, or $z$-axis, and the scattered neutrons to be parallel or antiparallel to the $x$-, $y$-, or $z$-axis, we can measure neutron spin-flip (SF) cross sections $\sigma^{\rm SF}_{x,y,z}$ (Fig. \ref{fig:polarised data}a) \cite{RevModPhys.87.855,PhysRevLett.111.107006,PhysRevB.85.214516,PhysRev.181.920}. 
Since SF neutron scattering $\sigma^{\rm SF}_{x,y,z}$
can only measure magnetic scattering perpendicular to $\mathbf{Q}$ and neutron spin direction, in the ideal case, $\sigma^{\rm SF}_{x}=M_y+M_z$, $\sigma^{\rm SF}_{y}=M_z$, and $\sigma^{\rm SF}_{z}=M_y$.
At $\mathbf{Q}=(0, 0.577, 0)$, we have $\sigma^{\rm SF}_{x}\propto M_y+M_z$, $\sigma^{\rm SF}_{y}\propto M_z=M_a$, $\sigma^{\rm SF}_{z}\propto M_y=M_c$.
Therefore, since $\sigma^{\rm SF}_{y}\approx \sigma^{\rm SF}_{z}$ at $\mathbf{Q}=(0, 0.577, 0)$, spin fluctuations within the $ac$ plane are isotropic without the need for any data manipulation. However, to conclusively determine $M_b$, one must take additional $\sigma^{\rm SF}_{x,y,z}$ data at $\mathbf{Q}=(0, 0.577, l)$ with $l=1,2,3$ to project $M_b$ into the perpendicular plane of $\mathbf{Q}$ \cite{RevModPhys.87.855,PhysRevLett.111.107006,PhysRevB.85.214516,PhysRev.181.920}. 
By comparing magnetic scattering at two or more $l$ values, one can determine the
values of $M_b$ \cite{PhysRevB.85.214516}.

The raw data of $\sigma^{\rm SF}_{x,y,z}(0, 0.577, l)$ with $l=0,1,2,3$ at $E=1.0$, 1.8, and 2.8 meV are shown with error bars, which represent the square roots of neutron counts according to the statistics, in Fig. \ref{fig:polarised data}d, e, f, respectively. Raw data of 
$\sigma^{\rm NSF}_{x,y,z}(0, 0.577, l)$
can be found in the supplementary material. 
In real polarised experiments, 
nuclear coherent scattering (NC), nuclear spin incoherent scattering (NSI), background scattering (BG), and imperfect neutron polarisation, measured
on nuclear Bragg peaks $(0,0,2)$ and $(0,1,1)$ as the
flipping ratio $R=\sigma^{\rm NSF}_{x}/\sigma^{\rm SF}_{x}\approx 46$, will modify the observed  $\sigma^{\rm SF}_{x,y,z}$ and its relationship with $M_{x,y,z}$ \cite{RevModPhys.87.855,PhysRevLett.111.107006,PhysRevB.85.214516,PhysRev.181.920} (see supplementary material). Nevertheless, we can conclusively solve $M_{x,y,z}$ without the need to measure the NC, NSI, BG separately. Since the raw data 
of $\sigma^{\rm SF}_{x,y,z}$ at all probed energies satisfies 
$\sigma^{\rm SF}_{x}>\sigma^{\rm SF}_{y}\approx \sigma^{\rm SF}_{z}$
at $\mathbf{Q}=(0, 0.577, 0)$ (Fig. \ref{fig:polarised data}d, e, f), we conclude 
$M_y\approx M_z\approx M_a\approx M_c\approx M_x/2$ without any further analysis.
Therefore, spin excitations are approximately isotropic within the $ac$ plane at all measured energies from $E=1$ to 2.8 meV.  When $l$ gradually moves away from 0 to 3 
at $\mathbf{Q}=(0, 0.577, l)$, we find $\sigma^{\rm SF}_{y}\rightarrow \sigma^{\rm SF}_{x}$ (Fig. \ref{fig:polarised data}g, h, i), meaning that the magnetic scattering measured by $\sigma^{\rm SF}_{z}$ decreases to 0 with increasing $l$ and $\theta$ angle (Fig. \ref{fig:polarised data}b).
Since $\sigma^{\rm SF}_{z}$ can only measure $M_y$ and the contribution from $M_b$ increases with increasing $l$, the decrease of $\sigma^{\rm SF}_z$ indicates the magnitude of $M_b$ is much smaller than $M_a$ and $M_c$ (Fig. \ref{fig:polarised data}b). Therefore, we conclude
that spin excitations in UTe$_2$ are highly isotropic within the $ac$ plane but have no measurable magnitude along the $b$-axis ($M_a\approx M_c$ and $M_b\approx 0$. This conclusion is robust for all measured energies and is true below and above $T_c$ (See supplementary information).

After decoupling the projection rule, we extract the spin excitations along the $y$ ($M_y$) and $z$-axes ($M_z$) (Fig. \ref{fig:polarised data}g-i) (see supplementary for details). The magnetic form factor of the U-dimer is drawn as the black line. A clear difference between $M_y$ and $M_z$ is shown in Fig. \ref{fig:polarised data}g-i. In $l=0$, $M_y\approx M_z$. As $l$ goes from 0 to 3, $M_y$ changes with the magnetic form factor, but $M_z$ gradually diminishes to 0, which agrees with our expectation of $M_a\approx M_c\gg M_b$. The signals at ${\bf Q}=(0, 0.577, l)$, where $l=0,1,2,3$, represent the same magnetic fluctuations, but their magnitude is different due to the combined effects of the magnetic form factor, projection rule, and resolution function \cite{PhysRevB.90.140502}. Combining them, we derive the ${\bf Q}$-independent dynamic spin correlations along the orthorhombic crystal axes, $\{M_a,M_b,M_c\}$ (see Fig. \ref{fig:polarised data}c). The spin excitations along the $a$-axis is approximately equal to that along the $c$-axis, and the fluctuations along the $b$-axis are zero within the error bar. The result agrees with our unpolarised data (Fig. \ref{fig:unpolarised}g), but conflicts with others' speculation \cite{PhysRevB.100.134502,doi:10.1126/science.aav8645}. The $b$-axis, therefore, is a hard axis for magnetic fluctuations. 

\section{Theoretical model}

\subsection{Duality of $5f$ electrons}
The dual localized/itinerant nature of $5f$ electrons in uranium heavy-fermion compounds has proved to be a challenge from the theoretical modelling perspective. The duality manifests under various experimental probes. For example, signatures of atomic-like multiplets show up in X-ray spectroscopy even in highly itinerant magnets of U$M_2$Si$_2$ ($M=$Pd, Ni, Ru, Fe) \cite{Amorese_2020}. In INS studies of UPt$_2$Si$_2$, on the other hand, conflicting signatures of itinerant magnetism as well as local moments are seen, e.g., no clearly dispersing spin-waves but purely transverse fluctuations in the ordered state \cite{Lee_2018}. In the context of UTe$_2$, recent high-field measurements found features in the metamagnetic transition (MMT) with remarkably different temperature dependencies, that could be attributed to the response of local and itinerant moments \cite{weinberger2026}. To complement this, Knight shift measurements found signatures of relocalization at low temperatures pointing to the coexistence of hybridization and localized moments \cite{Azari_2025}.
Many of the above phenomena can perhaps be captured by starting with a microscopic model that includes hybridizing (delocalized and contributing to the Fermi surface) as well as non-hybridizing (localized and not contributing to the Fermi surface) $f$-moments.
Within the two-fluid framework of Nakatsuji, Pines and Fisk \cite{Nakatsuji_2004}, the interplay between these two components has been suggested to be at the heart of quantum criticality and unconventional superconductivity of many heavy-fermion compounds \cite{Lonzarich_2017}.  We briefly mention some of the previous theoretical attempts at incorporating both itinerant and localized components, starting with the duality model of Miyake and Kuramoto \cite{MIYAKE_1991}, the work of Ono on mixed-valent systems \cite{Ono_1998}, the modelling of the Fermi surface of UPd$_2$Al$_3$ by keeping some of the $5f$ electrons as localized and some as hybridizing \cite{Fulde_2003}, and the more recent theoretical studies in the context of UTe$_2$ \cite{Thomas23}.

Recent X-ray spectroscopy measurements have highlighted the $5f^2$ configuration as a good starting point for describing the low-energy physics of UTe$_2$, perhaps with a small admixture of the $5f^3$ configuration that would add further delocalized $f$-electrons \cite{Christovam2024}. Our minimal model will therefore start with two stable $5f$-moments on each uranium atom. To capture their duality, we will assume that only one of these electrons will interact strongly with the conduction electrons, hybridize and participate in the Fermi surface -- we will label it the $f_1$-electron. We will assume that the other, the $f_2$-electron, will be weakly coupled and remain in the localized, unhybridized state down to zero temperature (the localized $5f_2$-moment is depicted in Fig. \ref{fig:crystal}a). The same spectroscopy study \cite{Christovam2024} identified the $6d_{3z-r^2}$ orbitals of the uranium and the $5p_y$ orbitals of the Tellerium-2 ions as participating in the Fermi surface. In addition, a recent minimal tight-binding model including these orbitals, as well as $5f$-electrons \cite{Eaton2024}, obtained an excellent fit with quantum oscillation measurements. Our minimal model will thus include the $5f_{1,2}$-moments of the uranium coupled to the $6d_{3z-r^2}$ and $5p_y$ orbitals. These key ingredients, together with the crystal structure of UTe$_2$, are depicted in Fig. \ref{fig:crystal}a. To avoid any bias, we are using the same tight-binding parameters as Ref. \cite{Eaton2024}, but we are incorporating the $5f$-electrons differently. These form stable moments and will couple via the following Kondo exchange with the nearest $6d_{3z-r^2}$ and $5p_y$ orbitals
 \begin{eqnarray}
     H_K = \sum_{a=1}^2 \sum_{i  \eta\eta'  } \mathbf{S}_{fa,\eta} (\mathbf{r}_{i} )\cdot \left( \frac{J_{ad}}{4}\mathbf{S}_{d,\eta} (\mathbf{r}_{i}) +J_{ap}\mathbf{S}_{p,\eta'} (\mathbf{r}_{i})+J_{ap}\mathbf{S}_{p,\eta'} (\mathbf{r}_{i}-\mathbf{a})
   \right),
   \nonumber\\
\end{eqnarray}
where $\mathbf{r}_i$ denotes the centres of the uranium dimers, $d$ denotes the $6d_{3z^2-r^2}$ orbital on the uranium, $p$ denotes the $5p_y$ orbital on the Te-2, $\eta=1,2$ indexes the two uranium ions belonging to each uranium dimer, and $\eta'=1,2$  indexes the two Te-2 sublattices  around each uranium dimer. Each $f$-moment couples to the nearest $6d_{3z^2-r^2}$ orbital and four nearest $5p_y$ orbitals. As already mentioned, only the interaction with the $f_1$-moment is strong enough to result in its virtual delocalization and enlargement of the Fermi volume that gives the flat heavy-fermion bands. The coupling to the $f_2$-moment, on the other hand, will generate RKKY interactions between them.  As per the Doniach phase diagram, this restricts the Kondo couplings to: $J_{a=1}/W \sim 1$ and $J_{a=2}/W \ll 1$, where $W$ is the bandwidth. 

\subsection{Low-energy model for the localized moments}
We can integrate out the itinerant $6d_{3z^2-r^2}$, $5p_y$ and $f_1$-electrons to obtain an effective low-energy model for the localized $f_2$-moments
\begin{align}
    H_{\rm eff} =& -\frac{1}{2}\sum_{\substack{i\eta j\eta' \gamma \\ i\neq j \; {\rm for} \;\eta=\eta'}}
    \mathcal{J}^{\gamma}_{\eta\eta'}(\mathbf{r}_{i}-\mathbf{r}_{j})S^{\gamma}_{f2,\eta} \left(\mathbf{r}_{i}\right)S_{f2,\eta'}^{\gamma}\left(\mathbf{r}_{j}\right),
\end{align}
where the isotropic (independent of $\gamma=\{x,y,z\}$) part of the RKKY exchange between the localized $f_2$-moments is given by
\begin{eqnarray}
   \frac{1}{J_{2d}^2} \mathcal{J}^{}_{\eta\eta'}(\mathbf{r}_{i}-\mathbf{r}_{j})=
\chi_{dd}^{\eta\eta'} (\mathbf {r}_{i}-\mathbf{r}_{j})
+\frac{J_{2p}}{J_{2d}} \chi_{pd}^{\eta\eta'}(\mathbf{r}_{i}-\mathbf{r}_{j}) + \frac{J_{2p}^2}{J_{2d}^2}\chi_{pp}^{\eta\eta'} (\mathbf{r}_i-\mathbf{r}_j), 
\end{eqnarray}
with $\chi$ measuring the various orbital contributions. The Fourier transform is defined as $\mathcal{J}_{\eta\eta'}(\mathbf{Q})=\sum_{i}\mathcal{J}_{\eta\eta'}(\mathbf{r}_i)e^{-i\mathbf{Q} \cdot \mathbf{r}_i}-\delta_{\eta\eta'}\mathcal{J}_{\eta\eta}(\mathbf{r}=\mathbf{0})$, $\mathcal{J}_{}(\mathbf{Q}):=\frac{1}{2}(\mathcal{J}_{11}(\mathbf{Q})+\mathcal{J}_{12}(\mathbf{Q}))$, and similarly for the $\chi$ contributions. Details of the theoretical calculation can be found in Sec. \ref{sec:details}.

\subsection{Dimerization of the localized moments}
 Since the hopping parameters are independently set by fits to quantum oscillation measurements, our microscopic theory has only two tunable dimensionless parameters: $J_{1d}/W$, which sets the hybridization between the itinerant $5f_1$-moment and conduction electrons and hence the resulting band structure, and $J_{2p}/J_{2d}$ which sets the form of the total RKKY exchange $\mathcal{J}(\mathbf{q})$, up to an overall constant prefactor. We expect $J_{1p}\lesssim J_{1d}$ and this gives no $f-p$ hybridization at mean-field level, see Sec. \ref{sec:details} for further details. There is a large region in the 2D-parameter space that gives good agreement with experimental observations. The results of the calculation for representative choices can be found in Fig. \ref{fig:rkky}. Firstly, we find that the RKKY exchange between the nearest-neighbour $5f_2$-moments on the two uranium ions that make up the dimer is FM and stronger than any other RKKY exchange (more than an order of magnitude for some parameters). The magnitude of this exchange is the splitting between the two dispersions shown in Fig. \ref{fig:rkky}(e), which give the RKKY energy in the case of the two dimer $5f_2$-moments being ferromagnetically aligned, $\mathcal{J}_{}(\mathbf{Q})=\frac{1}{2}(\mathcal{J}_{11}(\mathbf{Q})+\mathcal{J}_{12}(\mathbf{Q}))$, and in the case of them being antiferromagnetically aligned, $\mathcal{J}_{\rm -}(\mathbf{Q})=\frac{1}{2}(\mathcal{J}_{11}(\mathbf{Q})-\mathcal{J}_{12}(\mathbf{Q}))$. We will assume that the AFM configuration is effectively gapped out and can be neglected in the low-energy response. In each uranium dimer, we thus have two ferromagnetically glued $5f_2$-moments that respond in unison as a single magnetic unit. The dimer moments interact via the RKKY exchange $\mathcal{J}_{}(\mathbf{Q})$, which we discuss next.

 \subsection{Dimer interactions}

\begin{figure}[H]
   \centering
   \includegraphics[width=0.5\linewidth]{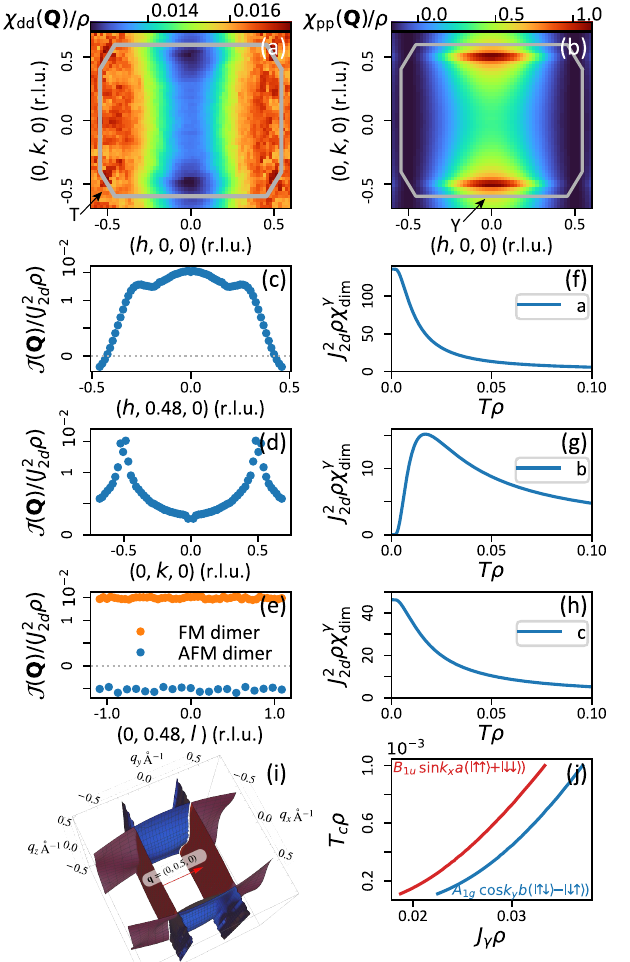}
   \caption{\textbf{Results of the dual model that includes localized and itinerant $5f$ electrons.} A representative choice of the two dimensionless couplings $J_{1d}\rho=2,J_{2p}/J_{2d}=1/8$ has been used, where $\rho$ is the density of states defined in Sec. \ref{sec:details}.
   \textbf{(a,b)} $5p_y$- and $6d_{3z^2-r^2}$- mediated contributions $\chi_{pp}(\mathbf{Q})$ and $\chi_{dd}(\mathbf{Q})$ to the inter-dimer RKKY exchange $\mathcal{J}(\mathbf{Q})=\frac{1}{2}(\mathcal{J}_{11}(\mathbf{Q})+\mathcal{J}_{12}(\mathbf{Q}))$ as a function of momentum $\mathbf{Q}$ in the $ab$-plane. Gray lines represent the Brillouin zone. For $\chi_{dd}(\mathbf{Q})$, $J_{1d}\rho=1$ was used. to capture the $T$-point resonance. \textbf{(c-e)} $\mathcal{J}(\mathbf{Q})$ scaled by $J_{2d}^2\rho$ along the 3 axes around the $\mathbf{Q}=(0, 0.48, 0)\;{\rm r.l.u.}$ maximum in the exchange. For the exchange between AFM dimers, we take the antibonding configuration $\mathcal{J}_-(\mathbf{Q})=\frac{1}{2}(\mathcal{J}_{11}(\mathbf{Q})-\mathcal{J}_{12}(\mathbf{Q}))$. The large energy splitting is a result of the strong FM exchange between the two $5f_2$-moments that form the uranium dimer. \textbf{(f-h)} The static susceptibility of an isolated dimer $\chi_{\rm dim}^{\gamma}$ along direction $\gamma$ for a representative choice of anisotropy parameters defined in the main text. At low temperatures, the susceptibility is quenched along the $b$-direction and tends to a constant for $a$ and $c$. \textbf{(i)} The Fermi surface with the quasi-2D cylindrical surfaces of mostly $6d_{3z^2-r^2}$ character (blue) or mostly $5p_y$ character (brown). The nesting $\mathbf{Q}$ vector between the two $5p_y$ character Fermi sheets is shown. \textbf{(j)} The superconducting critical temperature $T_c$ for the candidate states as a function of the pairing strength $J^{\gamma}$. $\gamma$ is $a$ and $b$ for the $B_{1u}$ and $A_{1g}$ states respectively. The $B_{3u}$ $\sin (k_zc/2)(|\uparrow\uparrow\rangle+|\downarrow\downarrow\rangle)$ triplet has $T_c\rho\gtrsim10^{-4}$ for $J_c\rho\gtrsim11$. (For the $A_{1g}$ state, we set $J_{a,c}=0$ for simplicity.)}
   \label{fig:rkky}
\end{figure}

As shown in Fig. \ref{fig:rkky}(i), the Fermi surface formed by the itinerant $5f_1$, $6d_{3z^2-r^2}$ and $5p_y$ orbitals has quasi-2D  character. There is a particularly strong nesting between the two sheets perpendicular to the $b$-direction with a nesting wavevector of $\mathbf{Q}=(0,0.48,0)$ {\rm r.l.u.} Figures \ref{fig:rkky}a and \ref{fig:rkky}b show the RKKY exchange mediated by the coupling of the localized $5f_2$-moments solely to the  $6d_{3z^2-r^2}$ and solely to the $5p_y$ itinerant orbitals, respectively for $J_{1d}\rho=1$. We can see that the $5p_y$-mediated exchange is enhanced in the vicinity of the $Y$ point, because of the well-nested Fermi sheets perpendicular to $b$ that have predominantly $5p_y$ character. The $6d_{3z^2-r^2}$-mediated exchange, on the other hand, leads to an enhancement of the exchange in the vicinity of the $T$ point. 
The maximum exchange mediated by the $5p_y$ orbitals is more than an order of magnitude stronger than that mediated by the $6d_{3z^2-r^2}$ orbitals and therefore controls the sharp $\mathbf{Q}$ features of the total RKKY exchange. We also find that the $T$-point resonance is a lot more sensitive to the value of the Kondo coupling than the $Y$-point resonance which appears for all tested Kondo couplings in the range $0.5\lesssim J_{1d}\rho \lesssim 3$ (the Fermi surface has been found to be qualitatively similar over this parameter range). This is understandable given that $f-p$ hybridization is self-consistently determined to be zero in our mean-field theory.

Panels c, d, and e of Fig. \ref{fig:rkky}, respectively, show the total RKKY exchange along the $a$, $b$ and $c$ directions away from the maximum in the vicinity of the $Y$ point, for a representative parameter choice. The dispersion is softer along the $a$-direction because the $b$-sheets have little curvature along $a$, and also because $\chi_{pp}(\mathbf{Q})$ and $\chi_{dd}(\mathbf{Q})$ grow in opposite directions along $a$. In summary, our calculations establish a clear hierarchy of RKKY exchange between the localized $5f_2$-moments: $\mathcal{J}_{12}(\mathbf{r}=\mathbf{0}) > \mathcal{J}_{\eta\eta}(\mathbf{e}_{a,b})$ (Fig. \ref{fig:crystal}a), where $\mathbf{e}_{\gamma}$ is the vector between n.n. uranium ions along direction $\gamma$. The intra-dimer exchange $\mathcal{J}_{12}(\mathbf{r}=\mathbf{0})$ between the n.n. $5f_2$-moments  along $c$ is greater than the {\it inter}-dimer exchange between the n.n. $5f_2$-moments along the $a$ or $b$ directions $\mathcal{J}_{\eta\eta}(\mathbf{e}_{a,b})$, which in turn is more than an order of magnitude stronger than RKKY exchange linking dimers belonging to different $ab$-planes. This RKKY exchange hierarchy, shown in Fig. \ref{fig:crystal}a is a generic consequence of the band structure. It holds over a large portion of our 2D parameter space -- we have verified it for $0.5\lesssim J_{1d} \rho\lesssim 3 $ and $0.2\lesssim J_{1p}/J_{1d}\lesssim 1 $.

\subsection{Magnetic anisotropy}
Because of SOC and crystal electric fields (CEF), the spin degrees of freedom of the $5f_{1,2}$-moments are endowed with an effective single-ion anisotropy, which most generally can be written as
\begin{eqnarray}
    H_{\rm anis}= \sum_{i \eta \gamma } D^{\gamma} S_{f1,\eta}^{\gamma}(\mathbf{r}_{i})S_{f2,\eta}^{\gamma}(\mathbf{r}_{i}),
    \label{eq:anis}
\end{eqnarray}
where $D^{\gamma}=\{D^x,D^y,D^z\}$ is the local orthorhombic anisotropy.
We find that to first order in $D^{\gamma}$ the RKKY exchange between spin components along direction $\gamma$ is enhanced/reduced across the whole Brillouin zone if $D^{\gamma}$ is positive/negative (details of the calculation can be found in sec. \ref{sec:details}). This is because the intra-dimer exchange -- the RKKY interaction between the nearest-neighbour $5f_2$-moments
\begin{align}
\mathcal{J}^{\gamma}_{12}(\mathbf{0})=\mathcal{J}^{}_{12}(\mathbf{0})+D^{\gamma}\left(J_{2d} \chi^{12}_{fd} (\mathbf{0}) + J_{2p}\chi_{fp}^{12} (\mathbf{0})\right),
\end{align}
experiences the greatest enhancement/reduction, and hence anisotropy is approximately of single-ion type, where the uranium dimer plays the role of the 'ion'. We have verified this behaviour over a wide region of our 2D parameter space: $0.75\lesssim J_{1d} \rho\lesssim 2 $ and $0.2\lesssim J_{1p}/J_{1d}\lesssim 1 $. From now on, we will thus neglect the anisotropy of the inter-dimer interactions. We have calculated the static susceptibility of an isolated dimer for $D^{\gamma}/(J_{2d}^2\rho)=\{ 0.088,-0.034,0.0076\}$, keeping the same relative ratios of anisotropy parameters as in Ref. \cite{Scott2026}, where the parameters were extracted through low-field fits. The results are shown in Fig. \ref{fig:rkky}(f-h), and we find that, generically, in the zero temperature limit, the lowest susceptibility $\chi^b_{\rm dim}$ drops to zero (i.e. along the hard $b$-axis with the lowest $D^{\gamma}$), whereas the other two tend to a constant value with $\chi^a_{\rm dim}>\chi^c_{\rm dim}$ for $D^a>D^c$.  Motivated by the INS results, we shall consider the limiting case, where $D^a=D^c$, and return to exploring the reasons for this emergent $O(2)$ symmetry later.

Models of strongly coupled pairs of spin$-1/2$ moments interacting with each other have been considered previously, e.g., by Tachiki and Yamada \cite{Tachiki_1970}, and have been found to harbour rich phase diagrams. Here,  for weak enough inter-dimer interactions and $D^{a}=D^c$, $D^b<0$, the spin$-1$ dimer moments have a non-magnetic $|0\rangle$ ground state and doubly-degenerate $|\pm1\rangle$ first-excited states with an energy gap $\Delta_1=\frac{1}{2}(\mathcal{J}^{a,c}_{12}(\mathbf{0})-\mathcal{J}_{12}^{b}(\mathbf{0}))$, where we have chosen the $b$-axis as the quantization axis. The low energy theory for the dimer moments has the following Hamiltonian
\begin{align}
    H_{O(2)} = \Delta_1\sum_i \left(S^z_{f2}(\mathbf{r}_i))\right)^2
    -\frac{1}{2}\sum_{i \neq j}\mathcal{J}(\mathbf{r}_i-\mathbf{r}_j)\left(S^x_{f2}(\mathbf{r}_i)S^x_{f2}(\mathbf{r}_j)
    +S^y_{f2}(\mathbf{r}_i)S^y_{f2}(\mathbf{r}_j)
    \right),
\end{align}
where $S_{f2}(\mathbf{r}_i)=\sum_{\eta}S_{f2,\eta}(\mathbf{r}_i)$ are the dimer moments. The gapped magnon excitations have the following spectrum 
\begin{align}
    \omega_{\rm mag}(\mathbf{Q})=\sqrt{\Delta_1^2-2\Delta_1\tilde{\mathcal{J}}_{}(\mathbf{Q})},
\end{align}
where $\tilde{\mathcal{J}}(\mathbf{{Q}})=\mathcal{J}(\mathbf{{Q}})-\frac{1}{2}\mathcal{J}_{12}(\mathbf{{r}=\mathbf{0}})$.
The gap closes at the quantum critical point (QCP) where the dimers order magnetically \cite{Pires_2008, PIRES2009}. Within the RPA approximation used here (details can be found in Sec. \ref{sec:details}), this takes place when ${\rm max}(\tilde{\mathcal{J}}(\mathbf{Q}))=\frac{\Delta_1}{2}$. The QCP belongs to the $O(2)$ (2D+1) universality class. We shall assume that the dimer interactions $\mathcal{J}(\mathbf{r}_i-\mathbf{r}_j)$ are sufficiently weak that we are still in the gapped phase, although the proximity to the QCP can serve to enhance $O(2)$ fluctuations of the dimer moments.

We propose an interesting scenario, where quantum critical fluctuations of the localized $5f_2$-moments coexist with the heavy Fermi liquid formed by the itinerant $5f_1$-moments and conduction electrons. As we approach the QCP, where the localized moments order, one can expect an associated instability of the itinerant $5f_1$-moments to relocalization. This association between quantum criticality and the fractions of the $f$-moments that hybridize and localize respectively has been explored previously in the 2-fluid-model framework that includes both \cite{Lonzarich_2017}. Indeed, signatures of relocalization have been found below $10$K in Knight shift measurements on UTe$_2$ \cite{Azari_2025}, signalling an approaching magnetic instability.

 \subsection{Inelastic response} 

Our minimal microscopic model characterizes the low-energy physics of UTe$_2$ as follows. Localized dimer moments interact via RKKY exchange that is mediated by the heavy-fermion bands formed from the itinerant $5f_1$, $5p_y$ and $6d_{3z^2-r^2}$ orbitals. An ac magnetic field leads to particle-hole excitations of the heavy-fermions and a dissipative response is generated via Landau damping. This dissipative response is amplified by the strong response of the localized dimer moments, particularly at the wavevectors where they are close to ordering magnetically. 
We can capture the resonance within an RPA-type approximation, where we write down the imaginary part of the susceptibility in the $ac$-plane as 
\begin{eqnarray}\label{paramagnon}
    \chi''^{ac}(\mathbf{Q},\omega)
    &=&\frac{\Gamma(\mathbf{Q})}{\tilde{\mathcal{J}}(\mathbf{Q})} 
    \frac{\omega}{\omega^2+ \left[\lambda(\mathbf{Q})+\frac{\omega^2}{\omega_c^2(\mathbf{Q})}\right]^2\Gamma(\mathbf{Q})^2 },\nonumber\\   \lambda(\mathbf{Q})&=& \left(\chi^{ac}_{\rm dim}\tilde{\mathcal{J}}(\mathbf{Q}) \right)^{-1}-1,
    \label{eq:chi''}
    \label{eq:res}
\end{eqnarray}
where $\Gamma(\mathbf{Q})$ is the Landau damping relaxation rate, $\omega_c(\mathbf{Q})$ sets the curvature of the reactive response, and we have expanded the dynamical RKKY interaction to quadratic order in frequency: $\tilde{\mathcal{J}}(\mathbf{Q},\omega)=\tilde{\mathcal{J}}(\mathbf{Q})(1+i\Gamma^{-1}(\mathbf{Q})\omega-\omega^2/\omega_c^2)$. We have assumed the overdamped $\Gamma(\mathbf{Q}) \ll \tilde{\mathcal{J}}(\mathbf{Q})$ limit, which allows us to approximate the dimer susceptibility with its static $\omega=0$ limit $\chi^{ac}_{\rm dim}=\frac{2}{\Delta_1}$. As we approach the QCP, where ${\rm max} (\tilde{\mathcal{J}}(\mathbf{Q}))=\chi^{ac}_{\rm dim}=\frac{2}{\Delta_1}$, and the magnon excitation gap closes, $\lambda(\mathbf{Q})\rightarrow0$ leads to an enhancement of the overdamped inelastic response. The maximum of $\chi''^{ac}(\mathbf{Q},\omega)$ at $\omega_{\rm max}(\mathbf{Q}) \sim \lambda(\mathbf{Q})\Gamma(\mathbf{Q}) $ scales as $1/(\lambda(\mathbf{Q})\tilde{\mathcal{J}}(\mathbf{Q}))$ with a width that scales as $1/\omega_{\rm max}(\mathbf{Q})$. Since $\chi^b_{\rm dim}=0$ and $\chi^a_{\rm dim}=\chi^c_{\rm dim}=\chi^{ac}_{\rm dim}$, there is no response for fields along $b$ and the susceptibility is isotropic in the $ac$-plane. See Sec. \ref{sec:details} for further details.

Fig. \ref{fig:unpolarised}f shows the unpolarised scattering intensity in the vicinity of the $Y$ point as a function of the energy transfer \cite{Duan2021}, fitted against the form in Eq. \ref{eq:chi''} and showing excellent agreement. Moving away from $\mathbf{Q}=(0,0.577,0)\ {\rm r.l.u}$ along the $a$-direction, we can perform such a fit for each wavevector to obtain the frequency $\omega_{\rm max}(\mathbf{Q})$ where the intensity is maximized, as well as the intensity at this point proportional to $\chi''(\mathbf{Q},\omega_{\rm max}(\mathbf{Q}))$. The theoretical dispersion $\omega_{\rm max}(\mathbf{Q})$ is plotted in Fig. \ref{fig:unpolarised}b and the corresponding maximum intensity in Fig. \ref{fig:unpolarised}e, again showing excellent agreement with experimental data. The data in Fig. \ref{fig:unpolarised}b is a combination of data of $E_i=3.32$ meV and 5.47 times the data of $E_i=12$ meV and symmetrized by a mirror plane at $[0,h,l]$ for better illustration. For fitting the dispersion (black dashed line in Fig. \ref{fig:unpolarised}b) and intensity (orange dots in Fig. \ref{fig:unpolarised}e), on the other hand, we use the raw data of $E_i=3.32$ meV, which explains the slight discrepancy between the fitted dispersion and the combined experimental data in panel (b). The jump in intensity around $h=0.25$ in Fig. \ref{fig:unpolarised}e could come from the instrument background noise.

 Fig. \ref{fig:unpolarised}d shows the INS intensity as a function of the wavevector, moving away from the resonance at the Y point in the $c$-direction \cite{Duan2021}.   
 The differences in spin fluctuation dispersions along
$b$ and $c$-directions represent differences in magnetic exchange couplings in these
two directions. 
The smearing of the $Y$ point intensity along the $c$ direction is consistent with $\mathcal{J}(\mathbf{Q})$ not varying along this direction, as shown in Fig. \ref{fig:rkky}e, due to the magnetic decoupling of dimer $ab$-planes.
 Previously, it was assumed that moments fluctuate predominantly along the $a$-direction \cite{PhysRevB.104.L100409}. This picture is summarised by Fig. \ref{fig:crystal}d. Since neutron scattering can only probe spin fluctuations perpendicular to the scattering wavevector $\mathbf{Q}$, 
 all the spin fluctuations along the $a$-axis would contribute at a $\mathbf{Q}$ in the $bc$-plane.   
 Moving along $c$, the magnetic scattering intensity should simply track the magnetic form factor $F^2(\mathbf{Q})$ of a single U dimer since the RKKY intra-dimer exchange $\mathcal{J}(\mathbf{Q})$ and hence $\chi''(\omega,\mathbf{Q})$ are approximately constant if we move along the $c$-axis.  If the magnetic scattering arises from coupled U-dimers, where
 the dimer fluctuations are near isotropic in the $ac$-plane, there would be a decreasing contribution from moments fluctuating along the $c$-direction as the wavevector rotates from the $b$-axis towards the $c$-axis. This gives rise to an additional factor of $\left(1-Q_z^2/2Q^2\right)$ and a much better fit to the magnetic scattering intensity (Fig. \ref{fig:unpolarised}g).
 The good agreement with the magnetic form factor of the dimer $F^2(\mathbf{Q})$ (see Sec. \ref{sec:details} for further details) confirms the picture in which the $5f_2$-moments sitting on the two uranium ions making up the dimer respond in unison because of the strong FM interaction between them.

 Using neutron polarisation analysis, we find 
 that there are no magnetic fluctuations along the $b$-direction to within the experimental error, whereas the fluctuations along $a$ and $c$ are the same to within the experimental 
 error (Fig. \ref{fig:polarised data}c). The full theoretical picture in which the dimer moments fluctuate in the $ac$-plane and are coupled by RKKY interactions that extend over the $ab$-plane is summarised in Fig. \ref{fig:crystal}c and e.

 \subsection{Emergent O(2) symmetry}
 The fact that the low-energy fluctuations of the dimer moments have an emergent O(2) symmetry in spite of the underlying orthorhombic lattice and strong SOC is remarkable and warrants attention. One possible reason is the strong incommensurability of the AFM fluctuations at $\mathbf{Q}_{\ast}=(0,0.577,0){\;\rm r.l.u.}$. In the absence of any in-$ac$-plane anisotropy, the low-energy fluctuations are helices with an axis along $b$ and moments rotating in the $ac$-plane.  The RKKY dispersion $\tilde{\mathcal{J}}(\mathbf{Q})$ is sharply peaked in the Brillouin zone along the $b$ direction which introduces an energy cost for stretching the helix to a commensurate configuration. The Hamiltonian describing the $T=0$ field theory of the magnetically ordered state of dimers when ${\rm max}(\tilde{\mathcal{J}}(\mathbf{Q}))>\Delta_1/2$ can be written as
 \begin{align}
 \mathcal{H}=\int \left(\frac{1}{2}\kappa\left(\partial_{\tau}\theta\right)^2+\frac{1}{2}\mathcal{J}_{xx}(\mathbf{Q_*})(\partial_x \theta)^2+\frac{1}{2}\mathcal{J}_{yy}(\mathbf{Q_*})(\partial_y \theta-\delta_c)^2 - t \cos \left(2\theta
 \right) \; \right) {\rm d} x {\rm d} y {\rm d}\tau,
 \end{align}
where the stiffness $\kappa>0$, $\theta(\mathbf{r},\tau)
$ gives the angle from the $a$ towards the $c$ axis of the dimer moment at position $\mathbf{r}$ and imaginary time $\tau$, the incommensurability $\delta_c=0.077{\;\rm r.l.u.}$ measures the distance to the nearest half-integer multiple of the reciprocal lattice vector, $\mathcal{J}_{xx}(\mathbf{Q}_*)$ and $\mathcal{J}_{yy}(\mathbf{Q}_*)$ give the second derivative of the RKKY energy at $\mathbf{Q_*}$ along $a$ and $b$ respectively, and $t=\frac{1}{2}(\mathcal{J}_{12}^{a}(\mathbf{0})-\mathcal{J}_{12}^{c}(\mathbf{0}))$ is the Ising in-$ac$-plane anisotropy that will be considered perturbatively. We can probe the instability of the commensurate Ising phase ($\theta=0$ or $\pi$) to forming incommensurate order by considering the energy cost (per unit area) of a soliton that involves a phase slip of $\pi$ along the $b$ direction. The phase is unstable to phase slip proliferation when the energy cost $\Delta E_{\rm cost}=4\sqrt{\mathcal{J}_{yy}(\mathbf{Q_*})t}-\pi\mathcal{J}_{yy}(\mathbf{Q_*})\delta_c<0$ \cite{Lazarides2009}, which takes place for strong enough commensurability $\delta_c>\frac{4}{\pi}\sqrt{\frac{t}{\mathcal{J}_{yy}(\mathbf{Q_*})}}$, and the transition is of Pokrovsky-Talapov type \cite{Pokrovsky}. If this criterion is satisfied, phase slips proliferate and $t\rightarrow 0$ in RG sense at large lengthscales -- the in-$ac$-plane anisotropy is washed out as we coarse-grain over the helical precession of the spins. In other words, we want the crossover lengthscale at which the O(2)-symmetry-breaking anisotropy becomes relevant to be much larger than the period of the helix: $\sqrt{t/\mathcal{J}_{yy}(\mathbf{Q_*})}\gg 1/\delta_c$. Of course, we need to be sufficiently close to the QCP such that the correlation length $\xi\gg1/\delta_c$. While we have used naive RG scaling here, it was found in a related model that the instability towards incommensurability is further enhanced by fluctuations \cite{Lazarides2009}.
We note that the proposed mechanism, which could be behind the emergent $O(2)$ symmetry of low-energy magnetic fluctuations in UTe$_2$, while a possibility in a local-moment picture, would not be possible in a purely itinerant picture where longitudinal fluctuations of the moments can stabilize incommensurate order purely along the easy-$a$-axis.

Because our model predicts a single-ion anisotropy, where the dimer plays the role of the ion, signatures of the emergent $O(2)$ symmetry should also be present in the uniform static susceptibility measurements. The anisotropy splits the RKKY exchange between the localized $5f_2$-moments that make up the dimer, and at high temperatures, generates a molecular field along direction $\gamma$ of $\lambda^{\gamma}=D^{\gamma}\left(J_{2d} \chi^{12}_{fd} (\mathbf{0}) + J_{2p}\chi_{fp}^{12} (\mathbf{0})\right)$.
This molecular field 'turns on' once $f-d,p$ hybridization takes place around the coherence temperature. Following Ref. \cite{Scott2026}, we can extract it from the total molecular field by only looking at the non-linear part of the inverse susceptibility, obtained after subtracting the high-temperature Curie-Weiss fit: $\chi^{-1}_{\rm anom}:=\chi^{-1}-\chi^{-1}_{\rm CW}$, such that $\lambda^\gamma-\lambda^{\gamma'}\approx1/\chi^{\gamma}_{\rm anom}-1/\chi^{\gamma'}_{\rm anom}$. Using the ambient pressure measurements of Ref. \cite{Aoki2021} gives $\Delta_1=\frac{1}{2}(\lambda^{b}-\lambda^{ac})\approx 42.5 \,{\rm mol}/{\rm emu}\sim 5.5 {\rm meV}$ and $t=\frac{1}{2}(\lambda^{b}-\lambda^{ac})\approx 1.25 \,{\rm mol}/{\rm emu}\sim 0.16 {\rm meV}$ in the low-temperature limit, where we have assumed an underlying spin-$1/2$ magnetic moment of $2\mu_B$ to estimate the corresponding energies. The low-temperature anisotropy between $a$ and $c$, parametrised by $t$, is more than 30 times smaller than the out-of-plane anisotropy, parametrised by $\Delta_1$, in agreement with INS data, and with incommensurability overcoming the in-$ac$-plane anisotropy of the intra-dimer exchange leading to an emergent $O(2)$ symmetry.  We note that the $D^{\gamma}$ parameters of Ref. \cite{Scott2026}, were extracted through fits to higher energy phenomena, such as the onset of coherence above $100$\,K, which we suspect are largely unaffected by the low-energy incommensurate intradimer fluctuations that we propose lead to an emergent $O(2)$ symmetry at lower temperatures.

Although the observed near-isotropy in the $ac$-plane could be a limitation of experimental resolution, the fact that the response of the system is otherwise vastly different along these two directions, but the observed difference in scattering intensity is so small, points to a deeper explanation, such as the criticality-driven emergent $O(2)$ symmetry that is proposed here. Uniform susceptibility along $a$ is around 4 times larger than along $c$ \cite{Aoki2021}, and the susceptibility maximum and metamagnetic jump both vanish when the field is rotated by only $25$ degrees from the $b$ axis towards $a$, but are still present when it is rotated by up to $50$ degrees towards $c$ \cite{Scott2026}. The differences in INS intensities between $a$ and $c$ components, on the other hand, cannot be resolved within the experimental error, but can certainly be resolved with respect to the $b$ component.
 The fact that the low-energy magnetic fluctuations are nearly isotropic in the $ac$-plane has important ramifications for superconducting pairing that these fluctuations can mediate, which we move to discussing next.

\subsection{Implications for superconducting pairing}

We now proceed to evaluating the various possible pairing channels that are mediated by the fluctuations of the localized $5f_2$-moments. We obtain the pairing interaction between the itinerant $5f_1$-moments
\begin{eqnarray}
H_{\rm pair} = -\frac{1}{2}\sum_{i ,j,\eta,\eta',\gamma} (D^{\gamma})^2 \chi_{\eta\eta'}^{\gamma}(\mathbf{r}_{i}-\mathbf{r}_{j})S_{f1,\eta}^{\gamma}(\mathbf{r}_{i})S_{f1,\eta'}^{\gamma}(\mathbf{r}_{j}), 
\label{eq:pair}
\end{eqnarray}
 where we have taken the zero-frequency limit
\begin{eqnarray}
 \chi_{\eta\eta' }^{\gamma}(\mathbf{r}_{i}-\mathbf{r}_{j})
 =\int_0^{\beta} \langle S_{f2,\eta}^{\gamma}(\mathbf{r}_{i},\tau)S_{f2,\eta'}^{\gamma}(\mathbf{r}_{j},0) \rangle \; {\rm d}\tau. 
\end{eqnarray}
For simplicity, we limit the pairing interactions to nearest neighbours along the three crystallographic directions, and we introduce an anisotropy parameter $\delta$ ($\delta$=0 for easy $ac$-plane anisotropy and $\delta=1$ for easy $a$-axis anisotropy)
\begin{align}
H_{\rm pair} = -\sum_{i ,\eta,\gamma} \left[J_{\gamma}(1+\delta) S_{f1,\eta}^{a}(\mathbf{r}_{i})S_{f1,\eta}^{a}(\mathbf{r}_{i}+\mathbf{e}_{\gamma}) +J_{\gamma}(1-\delta) S_{f1,\eta}^{c}(\mathbf{r}_{i})S_{f1,\eta}^{c}(\mathbf{r}_{i}+\mathbf{e}_\gamma)
\right], 
\end{align}
where $J_{\gamma}=\frac{1}{8}\sum_{\gamma'}(D^{\gamma'})^2\chi^{\gamma'}_{\eta\eta}(\mathbf{e}_{\gamma})$, $\mathbf{e}_{\gamma}$ is the vector to the nearest neighbour in the $\gamma$-direction, and for the sake of compactness, we have defined $S^{\gamma}
_{\eta}(\mathbf{r}_{i}+\mathbf{e}_c)$ equal to $S_{\bar{\eta}}^{\gamma}(\mathbf{r}_i)$ for $\eta=1$ and equal to zero for $\eta=2$.
Table \ref{table:SC} gives the pairing energy for the various competing superconducting states, normalised by the pairing amplitude that we can express in terms of the $5f_1$ fermionic annihilation operators $|D|^2=\sqrt{2}\sum_{\gamma\alpha\beta} |\langle f_{1,\eta\alpha}(\mathbf{r})f_{1,\eta\beta}(\mathbf{r}+\mathbf{e}_{\gamma}) \rangle |^2$, where again for the sake of compactness we have defined $f_{1,\eta\beta}(\mathbf{r}+\mathbf{e}_{c}):=f_{1,\bar{\eta}\beta}(\mathbf{r})$ for $\eta=1$ and zero for $\eta=2$. The energies are plotted as a function of the anisotropy parameter $\delta$ in Fig. \ref{fig:crystal}f. (The modulations are approximate and the full $\mathbf{k}$ dependence can be found in Sec. \ref{sec:details}. For the singlet state we simplified the situation by setting $J_{a,c}=0$ so that no admixtures are generated.)

 Triplet pairing has been suggested to be the most likely scenario for UTe$_2$ due to the critical field exceeding the Pauli limit and lack of significant Knight shift drops across $T_c$ \cite{Matsumura2023,Kitagawa2024,Matsumura2025} for all field directions.  However, strong AFM spin fluctuations of uranium moments and neutron spin resonance are observed at finite momentum transfer $\mathbf{Q}$ along the $b$ axis \cite{PhysRevLett.125.237003,PhysRevB.104.L100409,Butch2022UTe2,Duan2021,doi:10.7566/JPSJ.90.113706}. Upon application of a hydrostatic pressure, UTe$_2$ exhibits static AFM order at a similar wavevector \cite{PhysRevX.15.021075}, similar to all other magnetic-driven spin-singlet unconventional superconductors \cite{RevModPhys.84.1383}. Previous microscopic models of UTe$_2$'s superconductivity therefore took these AFM fluctuations as a strong candidate for superconducting pairing in the triplet channel by assuming Ising easy $a$-axis anisotropy, e.g. Ref.  \cite{Tei2024}.
From Fig. \ref{fig:crystal}f, we see that in the Ising limit ($\delta=1$), the pairing energy of the $A_{1g}$ $\cos k_yb(|\uparrow\downarrow\rangle- |\downarrow\uparrow\rangle)$ singlet is the same (normalized by the pairing amplitude) as that of the $B_{1u}$ $\sin k_yb(|\uparrow\downarrow\rangle+ |\downarrow\uparrow\rangle)$ triplet. Both have moments along $a$ and AFM pairing along the $b$-direction and are therefore stabilized by AFM exchange between the uranium ladders with easy $a$-axis anisotropy. The $B_{1u}$ $\sin k_yb(|\uparrow\downarrow\rangle+ |\downarrow\uparrow\rangle)$ triplet state would thus appear to be a possible candidate. Our results, however, show that this picture needs to be revised because spin fluctuations and neutron resonance around the $Y$ point are much closer to easy $ac$-plane anisotropy. As $\delta\rightarrow 0$ and approaching this limit, the pairing energy of the $B_{1u}$ $\sin k_yb(|\uparrow\downarrow\rangle+ |\downarrow\uparrow\rangle)$ triplet vanishes. Because of the emergent $O(2)$ symmetry, AFM exchange is no longer pair forming in the triplet channel but remains pair forming in the singlet channel. The above triplet state can no longer successfully compete against the singlet state and other candidates must be considered. 

Triplet states with a pairing energy that survives in the easy-plane $\delta=0$ limit must be formed by FM exchange with Cooper pair spins pinned to the $ac$-plane (dominant $\mathbf{d}$ vector component along $b$). These include the $B_{3u}$ $\sin (k_zc/2)(|\uparrow\uparrow\rangle+|\downarrow\downarrow\rangle)$ triplet driven by intradimer FM exchange, and the $B_{1u}$ $\sin k_x a(|\uparrow\uparrow\rangle+|\downarrow\downarrow\rangle)$ triplet, driven by FM exchange along the dimer ladders parallel to $a$. As shown in Fig. \ref{fig:crystal}f, the pairing energies of these two states do not evolve with the anisotropy parameter $\delta$. To analyse the competition between them as well as the proposed singlet state, we have calculated the critical temperature $T_c$ from the linearised gap equation in the easy-$ac$-plane limit ($\delta=0$).  Fig. \ref{fig:rkky}j shows $T_c$ for the candidate states as a function of the pairing strength $J^{\gamma}$. (When considering the singlet state, we set $J_{a,c}=0$ for simplicity, since this only increases its $T_c$.) The $B_{3u}$ $\sin (k_zc/2)(|\uparrow\uparrow\rangle+|\downarrow\downarrow\rangle)$ triplet requires at least a 100 times stronger pairing interaction to achieve $T_c\rho=10^{-4}$ than the proposed $B_{1u}$ triplet or the singlet, for all tested Kondo couplings in the range $0.5\lesssim J_{1d}\rho\lesssim3$. This is because it relies on the breaking of the inversion symmetry around the uranium ion and the associated matrix elements are small (see Sec. \ref{sec:details} for further details). This leaves the $B_{1u}$ state as the most likely triplet candidate, although the possibility of a singlet cannot be ruled out.
We also note that the $A_{u}$  $\sin(k_zc/2)(|\uparrow\uparrow\rangle -|\downarrow\downarrow\rangle)$ and the 
$B_{2u}$  $\sin(k_xa)(|\uparrow\uparrow\rangle -|\downarrow\downarrow\rangle)$ states, while stabilized by Ising fluctuations, have a vanishing pairing energy in the easy-$ac$-plane limit.

Although the general consensus points to  UTe$_2$ being a triplet superconductor, our results can be seen as increasing the possibility of it being in fact a singlet. The prominent magnetic fluctuations in the vicinity of the $Y$ point that were thought to be Ising are in fact easy-plane. This excludes the interladder AFM exchange, that is responsible for these $Y$-point fluctuations, from triplet pair formation. However, it can still stabilize singlet pairing. A crucial question that remains is how to reconcile a singlet state with the Pauli-exceeding upper critical fields that have been observed, in particular along the $b$-axis. Here again, our results offer an important perspective and point to strong SOC-induced anisotropy as a possible answer. The AFM fluctuations in the vicinity of the $Y$ point have a vanishing weight along the $b$-axis. The fluctuating dimers, made up of the localized $5f_2$-moments, are pinned strongly to the $ac$-plane and have a vanishing susceptibility along $b$, i.e., an effective Land{\' e} $g$-factor of zero along this direction. If we assume that the itinerant heavy-fermions are coupled more strongly to the fluctuating dimers than an external magnetic field then they could be protected from its effects until the fluctuations are altered in some way (we are essentially neglecting the direct Zeeman energy of the electrons). This in fact takes place at the metamagnetic transition when a sufficiently high field is applied along $b$ and the AFM fluctuations are quenched.

\begin{table}
\caption{Pairing energy of the competing superconducting states (normalized by the pairing amplitude). $J_{\gamma}$ is the pairing interaction strength for nearest neighbours along $\gamma$. The spin quantization axis is taken along $a$. $\delta$ is the anisotropy parameter with $\delta=0$ for the easy $ac$-plane limit, and $\delta=1$ for the easy $a$-axis limit.}\label{table:SC}
\begin{center}
\begin{tabular}{ccc}
Irrep.  & Pair state & $\langle H_{\rm pair}\rangle/|D|^2$ \\
\hline\hline
$B_{1u}$   & $\sin k_x a(|\uparrow\uparrow\rangle+|\downarrow\downarrow\rangle)$ & $-\frac{J_a}{4}$  \\
$A_{1g}$ & $ \cos k_y b(|\uparrow\downarrow\rangle-|\downarrow\uparrow\rangle)$ & $\frac{J_b}{4}$ \\
$B_{1u}$ & $ \sin k_y b(|\uparrow\downarrow\rangle +|\downarrow\uparrow\rangle)$ & $\frac{J_b\delta}{4}$ \\ 
$B_{3u}$ & $ \sin(k_zc/2)(|\uparrow\uparrow\rangle +|\downarrow\downarrow\rangle)$ & $-\frac{J_c}{4}$ \\ 
$A_{u}$ & $ \sin(k_zc/2)(|\uparrow\uparrow\rangle -|\downarrow\downarrow\rangle)$ & $-\frac{J_c\delta}{4}$ \\ 
$B_{2u}$ & $ \sin(k_xa)(|\uparrow\uparrow\rangle -|\downarrow\downarrow\rangle)$ & $-\frac{J_a\delta}{4}$ \\ 
\end{tabular}
\end{center}
\end{table}

Details of the calculations can be found in Sec. \ref{sec:further}, where we also make reference to previous microscopic models and look at the response of the proposed superconducting states to applied field, making contact with recent Knight shift experiments.

\section{Conclusion}

Our results revise the current microscopic picture of UTe$_2$'s superconductivity. With an easy $ac$-plane anisotropy, the AFM exchange behind the $Y$-point becomes unfavourable for the previously proposed triplet pairing channel, implying that, if superconductivity in UTe$_2$
is spin triplet, FM spin exchange provides a more natural pairing interaction.
In congruence with INS data, our theoretical calculations point to inter-dimer FM exchange as the glue that binds triplet Cooper pairs and highlight the dimerisation of uranium's moments as the central ingredient of uranium's superconductivity, magnetic fluctuations and anisotropy.

We have conclusively established that the fluctuating moments responsible for the observed magnetic scattering at the $Y$ point have a negligible component along the $b$-direction and are nearly isotropic in the $ac$-plane. There are three pieces of evidence that support this view: (i) The moment-direction resolved INS intensity at the $Y$ point is equal for the $a$ and $c$ components and vanishes for $b$ to within the experimental error; (ii) in the INS study, a much better fit is obtained for the unpolarised scattering intensity, moving away from the $Y$ point along the $c$ direction,  if the spins are assumed to be isotropically distributed in the $ac$-plane, rather than predominantly along the $a$-direction; (iii) the non Curie-Weiss contribution to anisotropy, that is greatly enhanced below the coherence temperature, is much weaker in the $ac$-plane than between the plane and the $b$-direction\cite{Scott2026}.
The fact that the AFM exchange behind the $Y$-point fluctuations is not Ising, with the easy-axis along $a$, as was previously often assumed given the uniform susceptibility measurements \cite{Aoki2021}, has profound implications for Cooper pair pairing mediated by spin fluctuations. With an easy $ac$-plane anisotropy, AFM exchange is no longer pair forming in the triplet channel. %
Our nearest-neighbour pairing interaction model points to the FM exchange along the uranium ladders as the most likely driver of spin triplet superconductivity at ambient pressure, giving rise to a $B_{1u}$ order parameter and a dominant $\mathbf{d}$-vector component in the $b$-direction.

We have been able to capture the main features of INS data within a single theoretical framework that includes $5f$ moments that contribute to the Fermi surface as well as those that do not and remain localized. This accounts for the dual itinerant-localized nature of uranium's $5f$ electrons.  The computed RKKY exchange between the localized moments gives rise to strong dimerization of uranium's spins; maxima near $Y$ and $T$ points of the Brillouin zone; and magnetic decoupling of dimers belonging to different $ab$-planes. Coupled with Landau damping of the heavy electrons, the maxima in the exchange between the localized moments can lead to the inelastic response observed at the $Y$ point.
We suggest that the response is further enhanced by proximity to a quantum critical point where strong incommensurability overwhelms Ising anisotropy giving rise to an emergent $O(2)$ symmetry. In the proposed picture, the near-critical local moments open a superconducting gap in the heavy Fermi liquid with which they coexist.

\section{Method}
\subsection{Synthesis}
Single crystals of UTe$_2$ were produced using the chemical vapor transport (CVT) method. Solid pieces of depleted uranium (99.98 \% purity) and tellurium (Sigma-Aldrich, 99.999 \% purity) were combined in a molar ratio of 2:3 and sealed with iodine (4 mg/cm$^3$, Sigma-Aldrich, 99.999 \% purity) in an evacuated quartz tube ($\ll 30$ mTorr). The quartz tubes had an inner diameter of 1.4 cm, an outer diameter of 1.8 cm, and a length of $\sim$19 cm. The tubes were placed in a single-zone furnace with the starting materials positioned at the hot end (furnace center) and held at 830 $^\circ$C for two weeks. The furnace was then cooled naturally to room temperature.

\subsection{Unpolarised Neutron experiment}
INS measurements on UTe$_2$ were carried out using the Cold Neutron
Chopper Spectrometer (CNCS) at Oak Ridge National Laboratory \cite{10.1063/1.3626935}. The crystals are naturally cleaved along the ab plane and form small flakes about 0.5–1 mm thick and up to 1 cm long. We co-aligned 27 pieces (total mass 0.9 g) of single crystals on oxygen-free Cu plates using an X-ray Laue machine to check the orientation of each single crystal\cite{Duan2021}. The crystal assembly is aligned in the $[H, K, 0]$ scattering plane and mounted on a $^3$He insert installed in the standard cryostat. The lowest temperature that can be reached in this setup is $T_\text{Base}$ = 0.25 K. INS data were collected with incident neutron energies set to $E_i = 12, 3.32$, and 2.5 meV. The data of the high-symmetry-cut is symmetrized to improve the statistics.

\subsection{Polarised Neutron experiment}
INS measurements on UTe$_2$ were carried out using the Three Axis Low Energy Spectrometer (ThALES) at the Institut Laue–Langevin (ILL). Around 111 pieces samples are co-aligned within the $[0,K,0]\times [0,0,L]$ plane on oxygen-free Cu plates using an X-ray Laue machine to check the orientation of each single crystal.
 
\subsection{Magnetic structure factor}
 The magnetic structure factor of the dimer moment is given by \cite{PhysRevB.104.L100409}
\begin{eqnarray}
    F(\mathbf{Q})=f(\mathbf{Q},U^{4+})\cos\left(Q_zd_\textbf{dimer} \right)
\end{eqnarray}
where $f(\mathbf{Q},U^{4+})$ is the magnetic form factor of U ion.
\begin{equation}
f(\mathbf{Q},U^{4+})=0.3291e^{-23.5475s^2}+1.0836e^{-8.454s^2}-0.434e^{-4.1196s^2}+0.0214
\end{equation}
where $s=|\mathbf{Q}|/4\pi$ in units of \AA${}^{-1}$. We only consider the $U^{4+}$ and $J_0$. Using $U^{3+}$ and $U^{5+}$ won't introduce too much difference. In this experimental range, $J_2$ and $J_4$ are almost constant and are small (See supplementary material). In Fig. \ref{fig:unpolarised}g, we assume the magnetic fluctuation has only a- and c-components with equal weights. The $\mathbf{Q}$ is at $(0,0.577,l)$, so the corresponding projection factors are $\hat{\mathbf{Q}}_b^2$ and $1$ respectively. The total projection factor normalized by $(0,0.577,0)$ is $1-Q_c^2/2Q^2$ since $Q_a$ is always to be 0.

\subsection{Theoretical calculation details}
\label{sec:details}
\label{sec:further}
We build on the tight-binding model of Ref. \cite{Eaton2024} which includes $5p_y$ and $6d_{3z^2-r^2}$ orbitals as well as fully itinerant $f$-electrons. We will keep the $5p_y$ and $6d_{3z^2-r^2}$ orbitals only and use exactly the same parameters for the hopping between them as well as their chemical potentials. We will then add a Kondo exchange term that couples the local $f_1$ and $f_2$ moments to the conduction electrons. The tight-binding Hamiltonian is given by
\begin{align}\nonumber
    H_t=&\epsilon_{p}\sum_{i \eta  } p^\dagger_{\eta\sigma}\left(\mathbf{r}_i \right) p_{\eta\sigma}\left(\mathbf{r}_i \right) +\epsilon_{d} \sum_{i \eta} d^\dagger_{\eta\sigma}\left(\mathbf{r}_i \right)d_{\eta\sigma}\left(\mathbf{r}_i \right)\\ \nonumber
    +&\frac{t^a_U}{2}\sum_{i \eta \pm} \left(d^\dagger_{\eta\sigma}\left(\mathbf{r}_i \right)d_{\eta\sigma}\left(\mathbf{r}_i  \pm\mathbf{a}\right)+{\rm h.c.}\right)\\ \nonumber
    +&t^c_U\sum_{i} \left(d^\dagger_{1\sigma}\left(\mathbf{r}_i\right)d_{2\sigma}\left(\mathbf{r}\right)+{\rm h.c.} \right)\\ \nonumber
    +&\frac{t^{p}_U}{2}\sum_{i,\boldsymbol{\alpha}\in A} \left(d^\dagger_{1\sigma}\left(\mathbf{r}_i \right) d_{2\sigma}\left(\mathbf{r}_i +\boldsymbol{\alpha}\right)+d^\dagger_{2\sigma}\left(\mathbf{r}_i \right) d_{1\sigma}\left(\mathbf{r}_i  -\boldsymbol{\alpha}\right)+{\rm h.c.} \right)\\ \nonumber
    +&t^{b}_T\sum_{i} \left(p^\dagger_{2\sigma}\left(\mathbf{r}_i \right)p_{1\sigma}\left(\mathbf{r}_i \right)+ p^\dagger_{2\sigma}\left(\mathbf{r}_i \right) p_{1\sigma}\left(\mathbf{r}_i  +\mathbf{b}\right)+{\rm h.c.} \right)\\ \nonumber
    +&\frac{t^{b}_{2T}}{2}\sum_{i,\eta,\pm}  \left(p^\dagger_{\eta\sigma}\left(\mathbf{r}_i \right)p_{\eta\sigma}\left(\mathbf{r}_i \pm \mathbf{b} \right)+{\rm h.c.} \right)\\
+&t_{UT}^{p}\sum_{i\eta\eta'} \left(d^\dagger_{\eta\sigma}\left(\mathbf{r}_i\right)p_{\eta'\sigma}\left(\mathbf{r}_i \right) +{\rm h.c.} \right)
\\ \nonumber
+&t_{UT}^{p}\sum_{i\eta\eta'} \left(d^\dagger_{\eta\sigma}\left(\mathbf{r}_i\right)p_{\eta'\sigma}\left(\mathbf{r}_i -\mathbf{a}\right) +{\rm h.c.}\right),
    \end{align}
where $p_{\eta\sigma}(\mathbf{r}_i)$ and $d_{\eta\sigma}(\mathbf{r}_i)$ are the fermionic annihilation operators for the $5p_y$ and $6d_{3z^2-r^2}$ orbitals respectively, $\mathbf{r}_i$ span the centres of the uranium dimers,
$A=\{ \mathbf{a}_2, \mathbf{a}_3,\mathbf{a}_2-\mathbf{a}_1,\mathbf{a}_3-\mathbf{a}_1\}$, $\mathbf{a}=a \,\hat{\mathbf{e}}_a$, $\mathbf{b}=b \,\hat{\mathbf{e}}_b$, $\mathbf{c}=c \,\hat{\mathbf{e}}_c$, 
$\{\mathbf{a}_1=\mathbf{a},\mathbf{a}_2=\frac{1}{2}\mathbf{c}+\frac{1}{2}\mathbf{b}+\frac{1}{2}\mathbf{a},\mathbf{a}_3=\frac{1}{2}\mathbf{c}-\frac{1}{2}\mathbf{b}+\frac{1}{2}\mathbf{a}\}$, and $\eta=1,2$ indexes the two U (or Te-2) sublattices.  The hopping parameters $t^{\gamma}_{X}$ describe the hopping in direction $\gamma$ between ions of $X$, with $p$ denoting the pyramidal links between uranium ions, or uranium and Te-2 ions. We define the approximate density of states as $\rho=1.025/\epsilon_d$, which simply sets the inverse energy scale, and allows us to define dimensionless couplings.

We can integrate out the delocalised $f_1$-moments as well as the $6d_{3z^2-r^2}$ and $5p_y$ conduction electrons to obtain an effective energy landscape for the $f_2$-moments. We do this by first decoupling the Kondo interaction between the band and the $f_1$-moments in the mean-field approximation
\begin{eqnarray}
    H_{\rm MF} =  -J_{1d}\sum_{i\eta\sigma} \left(f^\dagger_{1,\eta\sigma}\left(\mathbf{r}_{i}\right)d_{\eta\sigma}\left(\mathbf{r}_{i} \right)V_{fd} + {\rm h.c.} \right)
\nonumber\\
-J_{1p}\sum_{i\eta\sigma}\left(f^\dagger_{1,\eta\sigma}\left(\mathbf{r}_{i}\right)p_{\eta\sigma}\left(\mathbf{r}_{i} \right) V_{fp} + {\rm h.c.} \right), 
\end{eqnarray}
where the hybridisation fields $V_{fd}$ and $V_{fp}$ are self-consistently determined. We find that for $J_{1p} \lesssim J_{1d}$, which is very likely to be the case, the hybridisation field $V_{fp}=0$.
The full mean-field Hamiltonian can be written in matrix form as
\begin{equation}
    H=H_t+H_N=\sum_{\mathbf{k},\sigma}\Psi_{\sigma}^\dagger (\mathbf{k}) \begin{pmatrix}
        \lambda&0&-J_dV_{fd}&0&0&0\\
        0&\lambda&0&-J_dV_{fd}&0&0\\
        -J_dV_{fd}&0&\epsilon_d+\gamma_d&t^c_U+\frac{t_U^p}{2}f_d&\gamma_{}&\gamma_{}\\
        0&-J_dV_{fd}&t^c_U+\frac{t_U^p}{2}f^*_d&\epsilon_d+\gamma_d&\gamma_{}&\gamma_{}\\
        0&0&\gamma_{}^*&\gamma_{}^*&\epsilon_p+\gamma_p&t^b_T\\
        0&0&\gamma_{}^*&\gamma_{}^*&t^b_T&\epsilon_p+\gamma_p
    \end{pmatrix}
    \Psi_{\sigma}(\mathbf{k}),
\end{equation}
with $\Psi_{\sigma}(\mathbf{k})=(f_{1,1\sigma}(\mathbf{k}),f_{1,2\sigma}(\mathbf{k}),d_{1\sigma}(\mathbf{k}),d_{2\sigma}(\mathbf{k}),p_{1\sigma}(\mathbf{k}),p_{2\sigma}(\mathbf{k}))^T$ and $\gamma_d=t^a_U\cos(\mathbf{k}\cdot\mathbf{a})$,
$\gamma_p=t^b_{2T}\cos(\mathbf{k}\cdot \mathbf{b})$, $\gamma=2t_{UT}^p\cos(\mathbf{k}\cdot\mathbf{a}/2)$, $f_d=e^{-i \mathbf{k} \cdot   \mathbf{a}_2}+e^{-i \mathbf{k} \cdot   \mathbf{a}_3}+e^{i \mathbf{k} \cdot   (\mathbf{a}_1-\mathbf{a}_2)}+e^{i \mathbf{k} \cdot   (\mathbf{a}_1-\mathbf{a}_3)} $.
The mean-field equations that fix the hybridisation field $V_{fd}$ and the Lagrange multiplier $\lambda$ are given by
\begin{align}
    V_{fd}=\frac{1}{{\rm V_{\rm BZ}}}\sum_{\nu}\int_{\rm BZ}d\mathbf{k}\,U_{1\nu}(\mathbf{k})U_{\nu3}^\dagger(\mathbf{k})\frac{1}{1+\exp(\beta E_\nu(\mathbf{k}))},\\
    1=\frac{2}{\rm V_{\rm BZ}} \sum_{\nu}\int_{\rm BZ}d\mathbf{k}\,U_{1\nu}(\mathbf{k})U_{\nu 1}^\dagger(\mathbf{k})\frac{1}{1+\exp(\beta E_\nu(\mathbf{k}))},
\end{align}
where $U$ is the unitary matrix made up of the eigenvectors of the above Hamiltonian matrix in the columns and $\nu\in [1,6]$ indexes the eigenvectors. The orbital components of the Pauli susceptibility are given below
\begin{align}
   \chi_{\rm dd}^{\eta\eta'}(\mathbf{Q})&=\frac{1}{2V_{\rm BZ}}\sum_{\nu\nu'}\int_{\rm BZ}d\mathbf{k}U^*_{\eta+2,\nu}(\mathbf{k})U_{\eta'+2,\nu}(\mathbf{k})U^*_{\eta'+2,\nu'}(\mathbf{k}+\mathbf{Q})U_{\eta+2,\nu'}(\mathbf{k}+\mathbf{Q})\frac{n[E_{\nu'}(\mathbf{k}+\mathbf{Q})]-n[E_{\nu}(\mathbf{k})]}{E_{\nu}(\mathbf{k})-E_{\nu'}(\mathbf{k}+\mathbf{Q})}-\delta_{\eta\eta'}\chi_{\rm dd}^{\eta\eta'}(\mathbf{r}=\mathbf{0}),\nonumber\\
   \chi_{\rm pp}^{\eta\eta'}(\mathbf{Q})&=\frac{1}{2V_{\rm BZ}}\sum_{\nu\nu'\xi\xi'}\int_{\rm BZ}d\mathbf{k}U^*_{\xi+4,\nu}(\mathbf{k})U_{\xi'+4,\nu}(\mathbf{k})U^*_{\xi'+4,\nu'}(\mathbf{k}+\mathbf{Q})U_{\xi+4,\nu'}(\mathbf{k}+\mathbf{Q})\frac{n[E_{\nu'}(\mathbf{k}+\mathbf{Q})]-n[E_{\nu}(\mathbf{k})]}{E_{\nu}(\mathbf{k})-E_{\nu'}(\mathbf{k}+\mathbf{Q})}-\delta_{\eta\eta'}\chi_{\rm pp}^{\eta\eta'}(\mathbf{r}=\mathbf{0}),\nonumber\\\
      \chi_{\rm pd}^{\eta\eta'}(\mathbf{Q})&=\frac{1}{2V_{\rm BZ}}\sum_{\nu\nu'\xi}\int_{\rm BZ}d\mathbf{k}
      \frac{n[E_{\nu'}(\mathbf{k}+\mathbf{Q})]-n[E_{\nu}(\mathbf{k})]}{E_{\nu}(\mathbf{k})-E_{\nu'}(\mathbf{k}+\mathbf{Q})}(2\cos(\mathbf{k}\cdot\mathbf{a}/2))\nonumber\\
&[U^*_{\eta+2,\nu}(\mathbf{k})U_{\xi+4,\nu}(\mathbf{k})U^*_{\xi+4,\nu'}(\mathbf{k}+\mathbf{Q})U_{\eta+2,\nu'}(\mathbf{k}+\mathbf{Q})\nonumber\\
+
&U^*_{\xi+4,\nu}(\mathbf{k})U_{\eta'+2,\nu}(\mathbf{k})U^*_{\eta'+2,\nu'}(\mathbf{k}+\mathbf{Q})U_{\xi+4,\nu'}(\mathbf{k}+\mathbf{Q}) ]-\delta_{\eta\eta'}\chi_{\rm pd}^{\eta\eta'}(\mathbf{r}=\mathbf{0}), \nonumber\\
     \chi_{\rm fp}^{\eta\eta'}(\mathbf{Q})&=\frac{1}{2V_{\rm BZ}}\sum_{\nu\nu'\xi}\int_{\rm BZ}d\mathbf{k}
      \frac{n[E_{\nu'}(\mathbf{k}+\mathbf{Q})]-n[E_{\nu}(\mathbf{k})]}{E_{\nu}(\mathbf{k})-E_{\nu'}(\mathbf{k}+\mathbf{Q})}(2\cos(\mathbf{k}\cdot\mathbf{a}/2))\nonumber\\
&[U^*_{\xi+4,\nu}(\mathbf{k})U_{\eta',\nu}(\mathbf{k})U^*_{\eta',\nu'}(\mathbf{k}+\mathbf{Q})U_{\xi+4,\nu'}(\mathbf{k}+\mathbf{Q}) \nonumber\\
+
&U^*_{\eta,\nu}(\mathbf{k})U_{\xi+4,\nu}(\mathbf{k})U^*_{\xi+4,\nu'}(\mathbf{k}+\mathbf{Q})U_{\eta,\nu'}(\mathbf{k}+\mathbf{Q})]-\delta_{\eta\eta'}\chi_{\rm fp}^{\eta\eta'}(\mathbf{r}=\mathbf{0}),\nonumber\\
      \chi_{\rm fd}^{\eta\eta'}(\mathbf{Q})&=\frac{1}{2V_{\rm BZ}}\sum_{\nu\nu'}\int_{\rm BZ}d\mathbf{k}U^*_{\eta,\nu}(\mathbf{k})U_{\eta'+2,\nu}(\mathbf{k})U^*_{\eta'+2,\nu'}(\mathbf{k}+\mathbf{Q})U_{\eta,\nu'}(\mathbf{k}+\mathbf{Q})\frac{n[E_{\nu'}(\mathbf{k}+\mathbf{Q})]-n[E_{\nu}(\mathbf{k})]}{E_{\nu}(\mathbf{k})-E_{\nu'}(\mathbf{k}+\mathbf{Q})}-\delta_{\eta\eta'}\chi_{\rm fd}^{\eta\eta'}(\mathbf{r}=\mathbf{0}),
\end{align}
where in the case $\eta=\eta'$ we subtract the Brillouin zone average.
The general magnetic anisotropy of the $5f^2$ moment subspace is parametrized by Eq. \ref{eq:anis}. Treating the localized $5f_2$-moments in the static (classical) approximation, we can now integrate out the itinerant $5f_1$, $6d_{3z^2-r^2}$ and $5p_y$ electrons to obtain the following RKKY interaction, to quadratic order in $J_{2p}$ and $J_{2d}$, and to first order in the anisotropy parameter $D^{\gamma}$
 \begin{eqnarray}
   \frac{1}{J_{2d}^2} \mathcal{J}^{\gamma}_{\eta\eta'}(\mathbf{r}_{i}-\mathbf{r}_{j})=
\chi_{dd}^{\eta\eta'} (\mathbf{r}_{i}-\mathbf{r}_{j})
+\frac{J_{2p}}{J_{2d}} \chi_{pd}^{\eta\eta'}(\mathbf{r}_{i}-\mathbf{r}_{j}) + \frac{J_{2p}^2}{J_{2d}^2}\chi_{pp} (\mathbf{r}_i-\mathbf{r}_j) 
+ \frac{D^{\gamma}}{J_{2d}} \chi^{\eta\eta'}_{fd} (\mathbf{r}_i-\mathbf{r}_j) +\frac{D^{\gamma}J_{2p}}{J_{2d}^2}\chi_{fp}^{\eta\eta'} (\mathbf{r}_{i}-\mathbf{r}_{j}).\nonumber\\
\end{eqnarray}
Across a wide range of the two dimensionless couplings $0.75\lesssim J_{1d}\rho \lesssim 2$, $0.2\lesssim J_{2p}/J_{2d}\lesssim1$, we find that the larger the value of $D^{\gamma}$, the stronger the corresponding RKKY exchange $\mathcal{J}^{\gamma}(\mathbf{Q})$ for all wavevectors $\mathbf{Q}$. This is because anisotropy generates the greatest splitting for the strongest exchange, i.e., the FM intra-dimer exchange $\mathcal{J}^{\gamma}_{12}(\mathbf{r}=\mathbf{0})$. Anisotropy is thus of single-ion character, where the dimer plays the role of the ion.

Again, we have checked that for $0.5\lesssim J_{1d}\rho \lesssim 3$ and $0.2\lesssim J_{2p}/J_{2d}\lesssim1$, the intradimer RKKY exchange $\mathcal{J}^{\gamma}_{12}(\mathbf{r}=\mathbf{0})$ is the strongest and for many parameters more than an order of magnitude greater than any other exchange. This allows us to consider the dimers as the fundamental magnetic unit with an anisotropic susceptibility $\chi_{\rm dim}^{\gamma}(\omega)$. In our model, the two localized $5f_2$ spin-1/2 moments form a single spin-1 moment, i.e., the singlet can be projected out at low energies. With $D^{b} \lesssim D^{a,c}$, the dimer ground state and the first two excited states are given by
\begin{align}
    |0\rangle&=\frac{1}{\sqrt{2}}(f^{\dagger}_{2,1\uparrow}f^{\dagger}_{2,2\downarrow}+f^{\dagger}_{2,1\downarrow}f^{\dagger}_{2,2\downarrow})|{\rm vac}\rangle,
    \nonumber\\
    |\pm \rangle&=\frac{1}{\sqrt{2}}(f^{\dagger}_{2,1\uparrow}f^{\dagger}_{2,2\uparrow}\pm f^{\dagger}_{2,1\downarrow}f^{\dagger}_{2,2\downarrow})|{\rm vac}\rangle,
\end{align}
where $f^{\dagger}_{2,\eta\sigma}$ creates a $5f_2$-moment on the dimer sublattice $\eta$ with spin $\sigma$. The $b$-axis has been chosen as the quantisation axis.
To make contact with the emergent O(2) symmetry of the fluctuations, we set $D^{a}=D^c$, so that the ground states are degenerate with an excitation energy of $\Delta_1=\frac{1}{2}(\mathcal{J}^{a,c}_{12}(\mathbf{0})-\mathcal{J}_{12}^{b}(\mathbf{0}))=\frac{1}{2}(D^{a,c}-D^b)\left(J_{2d} \chi^{12}_{fd} (\mathbf{0}) + J_{2p}\chi_{fp}^{12} (\mathbf{0})\right)$, and $\chi^{a,c}_{\rm dim}(\omega)=\frac{2\Delta_1}{\Delta_1^2-(\omega+i\epsilon)^2}$, and $\chi^{b}_{\rm dim}(\omega)=0$.  We point out that the weak anisotropy $\Delta_1$ gives the dimers a vanishing effective moment , and as $\Delta_1\rightarrow 0$, a Curie response is recovered. Although this behaviour is specific to our particular model, an effective description in terms of localized dimer magnetic moments, governed by a particular CEF scheme, and interacting via RKKY exchange, is in general consistent with the observed INS data.

The dimers then interact with each other via the interdimer RKKY exchange $\tilde{\mathcal{J}}(\mathbf{Q})$. Making a Curie-Weiss mean-field approximation for the interdimer interactions, and applying an ac magnetic field $h^{\gamma}(\omega,\mathbf{Q})$ to the dimers, we can compute the expectation value of the magnetization
\begin{align}
    \sum_{\eta}\langle S_{f2,\eta}(\omega,\mathbf{Q})\rangle
    =\frac{\chi^{\gamma}_{\rm dim}(\omega)}{1-\chi^{\gamma}_{\rm dim}(\omega) \tilde{\mathcal{J}}(\mathbf{Q},\omega)}h(\omega,\mathbf{Q}),
\end{align}
where we have replaced the static RKKY exchange $\tilde{\mathcal{J}}(\mathbf{Q})$ with the dynamical one $\tilde{\mathcal{J}}(\mathbf{Q},\omega)$. We have neglected the itinerant contribution to the magnetization of the system (valid at weak-coupling $J_{2p,2d}\rho\ll1$). 
The dimers do not order magnetically as long as $\omega^2=\Delta_1^2-2\Delta_1\tilde{\mathcal{J}}(\mathbf{Q})>0$ for all $\mathbf{Q}$, i.e., the excitation spectrum is gapped everywhere. We assume that this condition is satisfied, although the system is likely close to criticality, giving a pronounced inelastic response. We note here that models of strongly coupled pairs of spin$-1/2$ moments interacting with each other have been considered previously, e.g., by Tachiki and Yamada \cite{Tachiki_1970}, and have been found to harbour rich phase diagrams.
The imaginary part of the susceptibility is given by
\begin{eqnarray}\label{paramagnon}
    \chi''^{\gamma}(\mathbf{Q},\omega)
    &=&\frac{\Gamma(\mathbf{Q})}{\tilde{\mathcal{J}}(\mathbf{Q})} 
    \frac{\omega}{\omega^2+ \left[\lambda(\mathbf{Q}) - \frac{\omega^2}{2\Delta_1\tilde{\mathcal{J}}(\mathbf{Q})} + \frac{\omega^2}{\omega_c^2(\mathbf{Q})}\right]^2\Gamma(\mathbf{Q})^2 },\nonumber\\
   \lambda(\mathbf{Q})&=& \left(\chi^{\gamma}_{\rm dim}(\omega=0)\tilde{\mathcal{J}}(\mathbf{Q}) \right)^{-1}-1,
    \nonumber\\
\end{eqnarray}
where we have expanded the dynamical RKKY exchange at small energies as $\tilde{\mathcal{J}}(\mathbf{Q},\omega)=\tilde{\mathcal{J}}(\mathbf{Q})(1+i\Gamma^{-1}(\mathbf{Q})\omega-\omega^2/\omega_c^2)$.
In the overdamped limit, where the relaxation rate $\Gamma(\mathbf{Q}) \ll \tilde{\mathcal{J}}(\mathbf{Q})$, the inelastic response $\chi''^{\gamma}(\mathbf{Q},\omega)$ is maximised at a resonant frequency  $\omega_{\rm res}(\mathbf{Q}) \sim \lambda(\mathbf{Q})\Gamma(\mathbf{Q}) $ that is much lower than the excitation energy $\sqrt{\Delta_1^2-2\Delta_1\tilde{\mathcal{J}}(\mathbf{Q})}$, which allows us to approximate the dimer susceptibility with its static limit $\chi^{\gamma}_{\rm dim}(\omega=0)$ and recover Eq. \ref{eq:res} in the main text.
The maximum of $\chi''^{\gamma}(\omega,\mathbf{Q})$ traces out the following dispersion
\begin{eqnarray}\nonumber
\omega_{\rm res}^2(\mathbf{Q})=\frac{\omega_c^4(\mathbf{Q})}{6\Gamma^2(\mathbf{Q})}
    &&\left[
\sqrt{\left(1+\frac{2\lambda(\mathbf{Q})\Gamma^2(\mathbf{Q})}{\omega_c^2(\mathbf{Q})}\right)^2+12\frac{\lambda^2(\mathbf{Q})\Gamma^4(\mathbf{Q})}{\omega_c^4(\mathbf{Q})}}
   -
   1-\frac{2\lambda(\mathbf{Q})\Gamma^2(\mathbf{Q})}{\omega_c^2(\mathbf{Q})} \right].
\end{eqnarray}

We now proceed to analysing the pairing interaction mediated by the fluctuations of the localized $5f_2$-moments, assuming easy $ac$-plane anisotropy, i.e., $D^{a}=D^{c}=D^{ac}$ and $D^b<0$ in the original model. Rewriting Eq. \ref{eq:pair} 
\begin{eqnarray}
H_{\rm pair} = -\frac{1}{2}\sum_{i \neq j,\eta,\eta',\gamma} (D^{\gamma})^2\chi_{\eta\eta'}^{\gamma}(\mathbf{r}_{i}-\mathbf{r}_{j})S_{f1,\eta}^{\gamma}(\mathbf{r}_{i})S_{f1,\eta'}^{\gamma}(\mathbf{r}_{j}) - \frac{1}{2} \sum_{i,\gamma,\eta\neq \eta'}
(D^{\gamma})^2
\chi^{\gamma}_{12}(\mathbf{0})S_{f1,\eta}^{\gamma}(\mathbf{r}_{i})S_{f1,\eta'}^{\gamma}(\mathbf{r}_{i}) ,
\end{eqnarray}
where we have also isolated the intra-dimer pairing interaction. The dimer has a vanishing susceptibility along $b$ at zero temperature, and we assume an emergent O(2) symmetry in the $ac$-plane, which leads to
\begin{eqnarray}
    \chi_{\eta\eta'}^{a}(\mathbf{r})=\chi_{\eta\eta'}^{c}(\mathbf{r}) =\chi_{\eta\eta'}^{ac}(\mathbf{r})\quad {\rm and} \quad  \chi_{\eta\eta'}^{b}(\mathbf{r})=0.
\end{eqnarray}
The INS data strongly support this above and below $T_c$, at least in the vicinity of the $Y$ point. Given that our theoretical model shows the anisotropy to be mainly of single-dimer character, we will assume that this approximation holds right across the Brillouin zone. 
With this approximation, the pairing interaction becomes (choosing the $b$-axis as the quantisation axis)
\begin{eqnarray}
    H_{\rm pair} = \frac{1}{4} (D^{ac})^2 \sum_{i \neq j,\eta,\eta'} \left(D^{\ast \downarrow\uparrow}_{\eta'\eta}(\mathbf{r}_j-\mathbf{r}_i) \chi_{\eta\eta'}^{ac}(\mathbf{r}_i-\mathbf{r}_j)f_{1,\eta \downarrow}(\mathbf{r}_i) f_{1,\eta' \uparrow}(\mathbf{r}_j) +{\rm h.c.} \right) 
    \nonumber\\
    + \frac{1}{4} (D^{ac})^2 \sum_{i,\eta\neq\eta'} \left(D^{\ast \downarrow\uparrow}_{\eta'\eta}(\mathbf{0}) \chi_{12}^{ac}(\mathbf{0})f_{1,\eta \downarrow}(\mathbf{r}_i) f_{1,\eta' \uparrow}(\mathbf{r}_i) +{\rm h.c.} \right),
    \nonumber\\
\end{eqnarray}
where have decoupled the interaction in the pairing channel, $D^{ac}=D^a=D^c$, and the anomalous pairing amplitude is given by
\begin{align}
   D^{\sigma\sigma'}_{\eta\eta'} (\mathbf{r}_i-\mathbf{r}_j)= \langle f_{1,\eta \sigma}(\mathbf{r}_i) f_{1,\eta' \sigma'}(\mathbf{r}_j)\rangle.
\end{align}
(The on-site term has been dropped. It enhances the on-site Coulomb repulsion, which favours triplet pairing.) 
Let us consider singlet and triplet pairing amplitudes, which satisfy
\begin{align}
    D^{\downarrow\uparrow}_{\eta\eta'} (\mathbf{r})= \zeta_{s,t}D^{\downarrow\uparrow}_{\eta'\eta} (-\mathbf{r}),
\end{align}
where $\zeta_{\rm s}=1$ for singlet pairing and $\zeta_{\rm t}=-1$ for triplet pairing.
Because of the inversion symmetry of the UTe$_2$ lattice, there is no coupling between these order parameters and the expectation value of the pairing interaction should be compared between the two cases
\begin{align}
    \langle H_{\rm pair} \rangle_{\rm s, t} &=\frac{1}{4} \zeta_{s,t}(D^{ac})^2 \sum_{i \neq j,\eta,\eta'}  |D^{\downarrow\uparrow}_{\eta\eta'}(\mathbf{r}_i-\mathbf{r}_j) |^2\chi_{\eta\eta'}^{ac}(\mathbf{r}_i-\mathbf{r}_j) +\frac{1}{4} \zeta_{s,t}(D^{ac})^2 \sum_{i ,\eta\neq\eta'}  |D^{\downarrow\uparrow}_{\eta\eta'}(\mathbf{0}) |^2\chi_{12}^{ac}(\mathbf{0})
    \nonumber\\
    &=\frac{1}{4N_s} \zeta_{s,t}(D^{ac})^2 \sum_{\mathbf{k},\mathbf{k}',\eta,\eta'}  D^{\ast\downarrow\uparrow}_{\eta\eta'}(\mathbf{k}) \left( \chi_{\eta\eta'
    }^{ac}(\mathbf{k}-\mathbf{k}') -\bar{\chi}^{ac}_{\eta\eta'}\right)D^{\downarrow\uparrow}_{\eta\eta'}(\mathbf{k'})  +\frac{1}{4} \zeta_{s,t}(D^{ac})^2 \sum_{i ,\eta\neq\eta'}  |D^{\downarrow\uparrow}_{\eta\eta'}(\mathbf{0}) |^2\chi_{12}^{ac}(\mathbf{0}),
\end{align}
where $\bar{\chi}^{ac}_{\eta\eta'}$ is the average of $\chi_{}^{ac}(\mathbf{k})$ over the Brillouin zone. We can thus see that FM exchange $\chi^{ac}_{\eta\eta'}(\mathbf{r})>0$ favours triplet pairing, whereas AFM exchange favours singlet pairing. AFM interactions incur an energy penalty in the case of triplet pairing.  Likewise, FM interactions incur an energy penalty in the case of singlet pairing. This should be contrasted with the case of Ising anisotropy where AFM interactions can be equally singlet as well as triplet pair forming (see below).

In the case of triplet pairing, the spin of the Cooper pair lies in the plane perpendicular to the $\mathbf{d}$ vector and we can write the triplet state as the following superposition 
\begin{align}
    \Psi(\mathbf{k}) = 
    d^z(\mathbf{k})(|\downarrow\downarrow\rangle  - |\uparrow\uparrow\rangle) + id^y(\mathbf{k}) (|\downarrow\downarrow\rangle +|\uparrow\uparrow\rangle)+
    d^x(\mathbf{k}) (|\uparrow\downarrow\rangle +|\downarrow\uparrow\rangle)
    \stackrel{{\rm easy} \; ac-{\rm plane}}{\approx}
    id^y(\mathbf{k}) (|\downarrow\downarrow\rangle +|\uparrow\uparrow\rangle),
 \end{align}
 where we have taken, as is the convention, $a$ as the quantization axis. We can see that easy-$ac$-plane anisotropy forces the Cooper pair spins to lie in the plane and the $\mathbf{d}$ vector has a dominant component along $b$.  Considering states with a $p$-wave dominant $\mathbf{d}$ vector component along $b$, the following three triplet states are possible (expanding to lowest order in $\mathbf{k}$): the fully gapped $A_u$ state with $\mathbf{d}(\mathbf{k})=(\eta_1k_x,k_y,\eta_3k_z)$, the $B_{1u}$ state with point nodes along the $c$-axis and $\mathbf{d}(\mathbf{k})=(\eta_1k_y,k_x,0)$, and the $B_{3u}$ state with point nodes along the $a$-axis and $\mathbf{d}(\mathbf{k})=(0,k_z,\eta_3k_y)$,
where $\eta_{1,2,3}$ parametrise small admixtures of the symmetry allowed, but subdominant, $\mathbf{d}$ vector components. (The $y$-component of the $\mathbf{d}$ vector of the $B_{2u}$ state is $f$-wave and this state has therefore been omitted from the analysis.)

It is insightful to take a simplified model, and limit the interactions in $\chi_{\eta\eta'}^{ac}(\mathbf{r}_i-\mathbf{r}_j)$ to nearest neighbours only along the three crystallographic axes so that 
\begin{align}
   \frac{1}{4}(D^{ac})^2 \chi_{\eta\eta'}^{ac}(\mathbf{Q})=2J_a \delta_{\eta\eta'} \cos q_x a+2J_b \delta_{\eta\eta'} \cos q_y b + J_c \delta_{\eta\bar{\eta}'},
\end{align}
where $J_{a,c}>0$ and $J_b<0$ for FM and AFM interactions respectively. Treating the superconductivity in the weak-coupling limit, we neglect any interband pairing and only consider the two bands that cross the Fermi level. We can then write the gap function as 
\begin{align}
    \Delta^{\downarrow\uparrow}(\mathbf{k})=\frac{\zeta_{s,t} (D^{ac})^2}{4N_s}\sum_{\mathbf{k}'\eta\eta'}u^*_\eta(-\mathbf{k})
    u^*_{\eta'}(\mathbf{k}) \chi_{\eta\eta'}^{ac}(\mathbf{k}-\mathbf{k}') u_\eta(-\mathbf{k}')
    u_{\eta'}(\mathbf{k}')
    \langle
    c_{\downarrow}(-\mathbf{k}')
     c_{\uparrow}(\mathbf{k}')
    \rangle,
\end{align}
where $c_{\sigma}(\mathbf{k})$ are the annihilation operators for the two bands crossing the Fermi surface (we have dropped the band index for simplicity) and $u_{\eta}(\mathbf{k})$ the matrix elements linking $f_{1,\sigma\eta}(\mathbf{k})$ to $c_{\sigma}(\mathbf{k})$.
The corresponding gap equation is given by
\begin{align}
    \Delta^{\downarrow\uparrow}(\mathbf{k})=-\frac{\zeta_{s,t}(D^{ac})^2}{4N_s}\sum_{\mathbf{k}'\eta\eta'}u^*_\eta(-\mathbf{k})
    u^*_{\eta'}(\mathbf{k}) \chi_{\eta\eta'}^{ac}(\mathbf{k}-\mathbf{k}') u_\eta(-\mathbf{k}')
    u_{\eta'}(\mathbf{k}')
    \frac{\tanh \left(
\frac{\sqrt{E({\mathbf{k}'})^2+|\Delta^{\downarrow\uparrow}(\mathbf{k}')|^2}}{2T}\right)}{2\sqrt{E({\mathbf{k}'})^2+|\Delta^{\downarrow\uparrow}(\mathbf{k}')|^2}} \Delta^{\downarrow\uparrow}(\mathbf{k}').
\end{align}
There are solutions for the $B_{3u}$ and $B_{1u}$ triplet states, which explore the FM intra-dimer and FM interactions along the uranium ladder, respectively. Note that, as expected, the $A_u$ triplet, with $\Delta^{\downarrow\uparrow}_{A_u}(\mathbf{k})=\Delta_{0}\sum_{\eta} u^*_{\eta}(\mathbf{k})u^*_{\eta}(\mathbf{-k})\sin k_yb$,  does not survive in the easy $ac$-plane limit and is not a solution of the above gap equation. For $B_{1u}$, we have
$\Delta^{\downarrow\uparrow}_{B_{1u}}(\mathbf{k})=\Delta_{0}\sum_{\eta} u^*_{\eta}(\mathbf{k})u^*_{\eta}(\mathbf{-k})\sin k_xa$, whereas for $B_{3u}$, we have $\Delta^{\downarrow\uparrow}_{B_{3u}}(\mathbf{k})=\Delta_{0}\sum_{\eta} \left( u^*_{\eta}(\mathbf{k})u^*_{\bar{\eta}}(\mathbf{-k}) -u ^*_{\eta}(-\mathbf{k})u^*_{\bar{\eta}}(\mathbf{k}) \right) $, which changes sign at $k_z=0$. We note that a singlet state of the form $\Delta^{\downarrow\uparrow}_{A_{1g}}(\mathbf{k})\approx\Delta_{0}\sum_{\eta} u^*_{\eta}(\mathbf{k})u^*_{\eta}(\mathbf{-k}) \cos k_yb$ is also possible, where we have neglected admixtures generated by non-zero $J_{a}$ and $J_{c}$. The state is stabilized by the AFM fluctuations between uranium ladders. We can obtain the corresponding $T_c$ by linearizing the gap equation in $\Delta_0$ and our results for a representative coupling are shown in Fig. \ref{fig:rkky}(j). The $B_{3u}$ triplet relies on the breaking of inversion symmetry with respect to the uranium ion and a non-zero value of the imaginary part of $u^*_{\eta}(\mathbf{k})u^*_{\bar{\eta}}(\mathbf{-k})$, which turns out to be small for the tested Kondo couplings in the range $0.5\lesssim J_{1d}\rho \lesssim 3$. The required pairing strength for $T_c\rho=10^{-4}$ is therefore 100 times stronger for the $B_{3u}$ triplet than the other two candidate states. Although a more detailed analysis of the fluctuations of the localized moments generated by the calculated RKKY exchange is necessary to precisely characterize the pairing strengths $J^{\gamma}$ and pairing interactions beyond nearest-neighbour should also be considered, this would point to the $B_{1u}$ state as the most likely candidate for triplet superconductivity. The possibility of a singlet, however, cannot be ruled out.

 \subsubsection{Previous models}

Our conclusions should be contrasted with the case of Ising anisotropy, which is what was previously thought to be the case for the $Y$ point fluctuations with $\chi_{\eta\eta'}^{a}(\mathbf{Q})\gg \chi_{\eta\eta'}^{b,c}(\mathbf{Q}) $. Choosing the $a$-axis as the quantisation axis, the Hamiltonian becomes
\begin{align}
    \langle H_{\rm pair} \rangle_{\rm Ising} =&-\frac{1}{8} (D^a)^2 \sum_{i\neq j,\eta,\eta',\sigma,\sigma'}{\rm sgn}(\sigma\sigma')  |D^{\sigma\sigma'}_{\eta\eta'}(\mathbf{r}_i-\mathbf{r}_j) |^2\chi_{\eta\eta'}^{a}(\mathbf{r}_i-\mathbf{r}_j) 
    \nonumber\\
    &-\frac{1}{8} (D^a)^2 \sum_{i ,\eta,\eta',\sigma,\sigma'}  {\rm sgn}(\sigma\sigma')|D^{\sigma\sigma'}_{\eta\eta'}(\mathbf{0}) |^2\chi_{12}^{a}(\mathbf{0}).
\end{align}
As in the case of easy $ac$-plane anisotropy, the direction of the $\mathbf{d}$ vector is not independent from the spatial structure. However, all three directions are possible in the case of Ising anisotropy, whereas the $b$ direction is strongly preferred in the case of easy $ac$-plane anisotropy. In particular, AFM interactions between the uranium ladders can be triplet pair forming for $\sigma=-\sigma'$, leading to a $B_{1u}$ triplet state with the $\mathbf{d}$ vector having a significant $a$-axis as well as well as $b$-axis component. The intra-dimer FM exchange can now lead to an $A_{u}$ triplet with the $\mathbf{d}$ vector pointing along $c$ as well as the $B_{3u}$ triplet with the $\mathbf{d}$ vector pointing along $b$. These were the conclusions reached in Ref. \cite{Tei2024}.

The theoretical model of Ref. \cite{Hakuno2024} found that at weak $fp$ hybridisation (which was determined to be zero in our mean-field theory), the spin-triplet $B_{1u}$ and $B_{3u}$ states are also stabilised by FM correlations along the one-dimensional U ladders, whereas at stronger $fp$ hybridisation the $B_{1g}$ and $B_{2g}$ spin-singlet states were stabilised by now AFM correlations along the one-dimensional U-ladders.

Finally, the model of Ref. \cite{Haruna2024} investigated superconductivity not mediated by spin-fluctuations but rather stemming from on-site Coulomb repulsion. Within a six orbital $f-d-p$ model, they reproduced the quasi-2D Fermi surface as well as the AFM fluctuations in the vicinity of the  $Y$ point. They found a highly anisotropic singlet state as the most probable candidate and a possible state at low-fields that would give a Knight shift drop along all directions.

\subsubsection{Pairing in the presence of an external field}
It is interesting to surmise how superconductivity is affected by the presence of a small external field. Although the preferred $\mathbf{d}$ vector has a dominant component along the $b$-direction, the strong pinning of the pair spin to the $ac$-plane will result in a drop in the susceptibility along all crystallographic directions, in agreement with the recent Knight shift measurements of Refs. \cite{Matsumura2023,Kitagawa2024,Matsumura2025}. 
Let us assume that the zero-field is the proposed $B_{1u}$ triplet with $\mathbf{d}(\mathbf{k})=(0,k_x,0)$. Applying a small field in the $a$-direction will take us to the $B_{1u}+iB_{2u}$ irrep with $\mathbf{d}(\mathbf{k})=(0,k_x,\alpha ik_x)$ and a small net spin in the $a$-direction proportional to $\alpha$. Applying a field in the $c$-direction will take us to the $B_{1u}+iA_{u}$ irrep with $\mathbf{d}(\mathbf{k})=(i\alpha k_x,k_x,0)$ and a small Cooper pair spin proportional to $\alpha$ in the $c$-direction. Finally, applying a small field in the $b$-direction will take us to the $B_{1u}+iB_{3u}$ irrep with $\mathbf{d}(\mathbf{k})=(i\alpha k_y,k_x,\alpha k_y)$ and a small Cooper pair spin proportional to $\alpha^2$.  However, there will be an energy penalty associated with the $\mathbf{d}$ vector rotation in all three cases, which will reduce the susceptibility below the normal state value (hence the Knight shift drop across $T_c$). We can estimate the characteristic field for which  the gain in the Zeeman energy outweighs this energy penalty, the Cooper pair spin is parallel to the applied field and the spin susceptibility returns to its normal state value. For the $a$ and $c$ directions we only lose a half of the condensation energy associated with the pairing of $c$ or $a$ spin components, respectively, whereas for the $c$ direction we lose the entire condensation energy once the spin is fully along $b$:
\begin{align}
    h^{*}_{a}=\frac{H_c/\sqrt{2}}{\sqrt{\Delta \chi_{a}}}\approx 1.3 \; {\rm T}
    \\
h^{*}_{c}=\frac{H_c/\sqrt{2}}{\sqrt{\Delta \chi_{c}}}
\approx 8.8 \; {\rm T}
    \\
    h^{*}_{b}=\frac{H_c}{\sqrt{\Delta \chi_{b}}}
    \approx 14.6 \; {\rm T}
\end{align}
using the values from Ref. \cite{Matsumura2023} for the $a$-direction and from Ref. \cite{Matsumura2025} for the $b$ and $c$ directions. These values are roughly consistent with the observed fields in Ref. \cite{Kitagawa2024} and Ref. \cite{Matsumura2025}. We note here that in the case of a field applied along $b$, all condensation energy is lost when we reach the Pauli limiting field $h^{*}_b$ since the Cooper pair spin has no component in the $ac$-plane. The only way for superconductivity to survive above $h^*_b$ is to transition to another state (see below).

\subsubsection{Hard to easy $b$-axis switch}

Ref. \cite{Scott2025} has shown that the $b$-axis becomes the hard-axis below a temperature that scales with the Kondo coupling. This coherence-driven mechanism collapses once a critical field is applied in the $b$-direction and the easy-axis reverts to the $b$-direction \cite{Scott2026}. We surmise that it also collapses at the critical pressure above which the system orders antiferromagnetically and the $b$-axis again becomes the easy axis \cite{PhysRevX.15.021075}. Thus, approaching the critical pressure or metamagnetic field drives us closer to the instability where the $b$-axis becomes the easy-axis. Approaching the transition at ambient pressure by increasing the field results in growing longitudinal fluctuations along the $b$-direction \cite{Tokunaga2023}. It is believed that these fluctuations drive a transition to the high-field SC phase above $h^*_b$. We surmise that such fluctuations, now associated with proximity to the critical pressure, could also drive the transition into the SC2 phase.
It is therefore tempting to extrapolate our results to the regime, where the $b$-axis becomes the easy-axis. Assuming that the other features of the RKKY exchange remain the same, we can propose the following triplet order parameters: 
\begin{itemize}
    \item the fully gapped $A_u$ phase with  $\mathbf{d}(\mathbf{k})=( \eta_1 k_x,\eta_2 k_y,\eta_3 k_z)$ with intra-dimer FM correlations, FM correlations along the ladders and AFM correlations between the ladders,
    \item the $B_{2u}$ phase with  $\mathbf{d}(\mathbf{k})=( \eta_1 k_z,0,\eta_3 k_x)$ with point nodes along the $b$-axis and FM intra-dimer correlations and FM correlations along the uranium ladders,
    \item The high-field phase could then belong to the fully gapped $A_u+iB_{2u}$ phase.
\end{itemize}
We can thus see that the growing longitudinal fluctuations could drive a transition from $B_{1u}+iB_{3u}$ with point nodes along the $c$-axis at low fields to the fully gapped $A_u+iB_{2u}$ at fields above $h^*_b$.

\bibliography{main}
\section*{Acknowledgements}
We thank Sheng Ran for providing some samples of UTe$_2$ and Qimiao Si for discussions. 
The neutron scattering experiments at Rice were supported by the U.S. DOE, BES under Grant Nos. DE-SC0026179 (P.D.). 
Part of the materials characterization work at Rice is supported by the Robert A. Welch Foundation under Grant No. C-1839 (P. D.).
Research at the University of California, San Diego was supported by the U.S. DOE, BES, under Grant No. DE FG02-04-ER46105 (single crystal growth) and the National Nuclear Security Administration under the Stewardship Science Academic Alliance Program through the U.S. DOE under Grant DE-NA0004235 (sample characterization). RB acknowledges support from the University of California, Santa Cruz through startup funds and from the National High Magnetic Field Laboratory, which is supported by the U.S. NSF through Grant No. DMR-1644779 and the state of Florida.

\section*{Author contributions statement}
P.D. conceived the project. Theoretical work is performed by E.S. and M.P.K.. Polarised neutron scattering experiments were carried out by D.W.T., P.B, P.S., and A.H, in discussion with P.D. and Z.W..   
Unpolarised neutron scattering experiments were carried out by C.D. and A.P., in discussion with P.D..
Single crystals are aligned by Z.W.  UTe$_2$ single crystals are grown by K.K.F., T.M.W., R.E.B, and M.B.M. The manuscript is written by M.P.K., P.D., Z.W., and E.S..
Data analysis is done by Z.W. and E.S. under the instruction of P.D. and M.P.K..
All authors reviewed the manuscript.

\section*{Additional information}

To include, in this order: \textbf{Accession codes} (where applicable); \textbf{Competing interests} (mandatory statement). 

The corresponding author is responsible for submitting a \href{http://www.nature.com/srep/policies/index.html#competing}{competing interests statement} on behalf of all authors of the paper. This statement must be included in the submitted article file.

\end{document}